\documentclass[aps,prl,twocolumn,superscriptaddress,floatfix,nofootinbib,showpacs,longbibliography]{revtex4-2}

\usepackage[utf8]{inputenc}
\usepackage[T1]{fontenc}     
\usepackage[british]{babel}  
\usepackage{enumitem}
\usepackage[sc,osf]{mathpazo}
\usepackage{times}
\usepackage[table]{xcolor}
\usepackage[scaled=0.86]{berasans}  
\usepackage[colorlinks=true, citecolor=purple, urlcolor=blue]{hyperref}
\usepackage{comment}
\makeatletter
\newcommand{\setword}[2]{%
  \phantomsection
  #1\def\@currentlabel{\unexpanded{#1}}\label{#2}%
}
\makeatother
\usepackage{comment}
\usepackage{graphicx} 
\usepackage[babel]{microtype}  
\usepackage{amsmath,amssymb,amsthm,bm,amsfonts,mathrsfs,bbm} 

\usepackage{xspace}  
\usepackage{pgf,tikz}
\usepackage{xcolor}
\usepackage{multirow}
\usepackage{array}
\usepackage{bigstrut}
\usepackage{braket}
\usepackage{color}
\usepackage{natbib}
\usepackage{multirow}
\usepackage{mathtools}
\usepackage{float}
\usepackage[caption = false]{subfig}
\usepackage{xcolor,colortbl}
\usepackage{color}
\newcommand{\Tr}{\operatorname{Tr}}

\newcommand{\be}{\begin{equation}}
\newcommand{\ee}{\end{equation}}
\newcommand{\ba}{\begin{eqnarray}}
\newcommand{\ea}{\end{eqnarray}}

\newcommand{\tr}{\operatorname{Tr}}

\newtheorem{theorem}{Theorem}
\newtheorem{corollary}{Corollary}
\newtheorem{definition}{Definition}
\newtheorem{proposition}{Proposition}
\newtheorem{observation}{Observation}

\def\>{\rangle}
\def\<{\langle}

\usepackage{centernot}
\usepackage{subfig}
\usepackage{filecontents}

\providecommand{\ket}[1]{| #1{\rangle}}
\providecommand{\bra}[1]{\langle #1|}

\providecommand{\id}{\mathbf{1}}
\providecommand{\dacb}[2]{\{\hspace{-.12cm}\{#1#2\}\hspace{-.12cm}\}}
\providecommand{\dcb}[2]{[\hspace{-.05cm}[#1#2]\hspace{-.05cm}]}

\usepackage[table]{xcolor}

\newtheorem{manualtheoreminner}{Theorem}
\newenvironment{manualtheorem}[1]{%
  \IfBlankTF{#1}
    {}
    {\renewcommand\themanualtheoreminner{#1}}%
  \manualtheoreminner
}{\endmanualtheoreminner}

\newtheorem{manualpropositioninner}{Proposition}
\newenvironment{manualproposition}[1]{%
  \IfBlankTF{#1}
    {}
    {\renewcommand\themanualpropositioninner{#1}}%
  \manualpropositioninner
}{\endmanualpropositioninner}

\newtheorem{manuallemmainner}{Lemma}
\newenvironment{manuallemma}[1]{%
  \IfBlankTF{#1}
    {}
    {\renewcommand\themanuallemmainner{#1}}%
  \manuallemmainner
}{\endmanuallemmainner}

\newtheorem{manualdefinitioninner}{Definition}
\newenvironment{manualdefinition}[1]{%
  \IfBlankTF{#1}
    {}
    {\renewcommand\themanualdefinitioninner{#1}}%
  \manualdefinitioninner
}{\endmanualdefinitioninner}

\newtheorem{manualremarkinner}{Remark}
\newenvironment{manualremark}[1]{%
  \IfBlankTF{#1}
    {}
    {\renewcommand\themanualremarkinner{#1}}%
  \manualremarkinner
}{\endmanualremarkinner}

\begin{document}

\title{Quantum Incompatibility in Parallel versus Antiparallel Spins}

\author{Ram Krishna Patra}
\affiliation{Department of Physics of Complex Systems, S. N. Bose National Center for Basic Sciences, Block JD, Sector III, Salt Lake, Kolkata 700106, India.}
\affiliation{HUN-REN Institute for Nuclear Research, P.O. Box 51, H-4001 Debrecen, Hungary.}

\author{Kunika Agarwal}
\affiliation{Department of Physics of Complex Systems, S. N. Bose National Center for Basic Sciences, Block JD, Sector III, Salt Lake, Kolkata 700106, India.}

\author{Biswajit Paul}
\affiliation{Department Of Mathematics, Balagarh Bijoy Krishna Mahavidyalaya, Balagarh, Hooghly-712501, West Bengal, India.}

\author{Snehasish Roy Chowdhury}
\affiliation{Physics and Applied Mathematics Unit, 203 B.T. Road Indian Statistical Institute Kolkata, 700108, India.}

\author{Sahil Gopalkrishna Naik}
\affiliation{Department of Physics of Complex Systems, S. N. Bose National Center for Basic Sciences, Block JD, Sector III, Salt Lake, Kolkata 700106, India.}

\author{Manik Banik}
\affiliation{Department of Physics of Complex Systems, S. N. Bose National Center for Basic Sciences, Block JD, Sector III, Salt Lake, Kolkata 700106, India.}

\begin{abstract}
We explore the joint measurability of incompatible qubit observables on ensembles of parallel and antiparallel spin-½ pairs. In parallel configuration, both spins are prepared in the same state, whereas in antiparallel case, each spin is paired with its flipped counterpart. We show that the antiparallel configuration uniquely enables  exact simultaneous prediction of three mutually orthogonal spin components—an advantage not achievable with parallel states. Extending beyond three observables, we examine joint measurability for larger sets of spin measurements and further generalize our analysis to state configurations beyond the parallel and antiparallel cases. As we show, our results reveal a deep connection to the `mean King retrodiction task' proposed by Vaidman, Aharonov, and Albert, and have implications for a cryptographic protocol introduced by Jeffrey Bub. We further demonstrate how the enhanced compatibility in the antiparallel configuration can facilitate efficient estimation of unknown measurement devices. Finally, we discuss prospects for experimentally realizing the enhanced measurement compatibility in antiparallel configuration by analyzing the effect on finite sub-ensembles of states.
\end{abstract}


\maketitle	
{\it Introduction.--} Bohr’s complementarity principle, a cornerstone of quantum theory, imposes fundamental limitation on simultaneous measurement of certain observables \cite{Bohr1928}. This is famously exemplified by the trade-off between path information and interference visibility in the double-slit experiment \cite{Wootters1979,Scully1989,Scully1991}, as well as by the impossibility of jointly measuring non-commuting observables such as position and momentum, or spin components along different axes \cite{Davies1976,Lahti1980,Busch1985}. Development of generalized measurements, formalized via positive operator-valued measures (POVMs) \cite{Kraus1983}, refined this understanding by demonstrating that incompatible observables can, in fact, be jointly measured—albeit with an inherent degree of fuzziness or imprecision \cite{Mittelstaedt1987,Busch1996,Heinosaari2016,Guhne2023}. Lately, measurement incompatibility has been shown to be intimately connected to other nonclassical phenomena, such as Bell nonlocality and Einstein-Podolsky-Rosen steering \cite{Wolf2009,Banik2013,Busch2013,Brunner2014,Quintino2014,Uola2014,Uola2015,Banik2015,Kar2016,Uola2020}. Measurement incompatibility also plays a pivotal role in several quantum protocols \cite{Guhne2023}, and thus motivates extensive investigations into the structure of incompatibility, including settings involving multiple copies of quantum systems \cite{Carmeli2016}.

In this Letter, we investigate whether the state configuration in a multi-copy setup offers any advantage in joint measurability. Focusing on the two-copy case, we compare two configurations: the parallel configuration of two identically prepared spin-½ particles (qubits) and the antiparallel configuration of a qubit paired with its spin-flipped counterpart. We demonstrate that a measurement device operating on antiparallel spin pairs can reproduce the statistics of spin measurements along the $x$, $y$, and $z$ directions—outperforming what is possible in the parallel case. Furthermore, we show that perfect joint measurability extends to more than three incompatible spin observables in the antiparallel setting. Moving beyond these two canonical cases, we consider more general configuration of the form $\rho_{\vec{m}} \otimes \Lambda(\rho_{\vec{m}})$, where $\Lambda$ is a positive trace-preserving (PTP) map \cite{Stinespring1955,Wigner1959,Jamiokowski1972,Choi1975}. Our construction also reveals a deep connection to other foundational concepts, such as the mean King retrodiction task \cite{Vaidman1987}, and enables improved robustness in the quantum key distribution protocol proposed by Bub \cite{Bub2001}. As we explore the perfect joint measurability of mutually unbiased qubit bases in the antiparallel configuration suggests a resource-efficient strategy for estimating unknown measurement devices. Finally, we discuss potential pathways toward experimental realization of our scheme by analyzing the notion of measurement compatibility on finite sub-ensembles of states.

{\it Measurement compatibility of unsharp spin observables.--} The state of a spin-½ particle is described by a qubit density operator $\rho_{\vec{m}} = \tfrac{1}{2}(\id + \vec{m} \cdot \vec{\sigma}) \in \mathcal{D}(\mathbb{C}^2)$, where $\vec{m} \in \mathbb{R}^3$ with $|\vec{m}| \leq 1$, $\vec{m} \cdot \vec{\sigma} := m_x \sigma_x + m_y \sigma_y + m_z \sigma_z$, the $\sigma$'s are the Pauli operators, and $\id$ is the identity operator. The state is pure when $|\vec{m}| = 1$, and mixed otherwise. An unsharp spin measurement along $\hat{n}$ is defined as $\sigma_{\hat{n}}(\lambda) \equiv \{\mathrm{P}^a_{\hat{n}}(\lambda) = \tfrac{1}{2}(\id + \lambda\, a\, \hat{n} \cdot \vec{\sigma})~|~a=\pm1\}$, where $\lambda \in [0,1]$ quantifies the sharpness of the measurement: $\lambda = 1$ corresponds to a projective (sharp) measurement, while $\lambda = 0$ corresponds to a completely uninformative (i.e., random-guess) measurement \cite{Busch1986}. More generally, a positive operator-valued measure (POVM) $\mathcal{M}$ is a collection of effects summing to the identity, i.e., $\mathcal{M} \equiv \{\pi_k \in \mathcal{E}(\mathbb{C}^2)~|~\sum_k \pi_k = \id\}$ \cite{Self1}. This framework allows us to define joint measurability of spin observables.
\begin{definition}\label{def1}
[Busch et al. \cite{Busch1996}] A set of spin observables $S \equiv \{\sigma_{\hat{n}_r}(\lambda)\}_r$ is said to be jointly measurable if there exists a POVM $\mathcal{G} \equiv \{\pi_{{\bf a}}\}_{{\bf a}}$ such that $\mathrm{P}^{a_r}_{\hat{n}_r}(\lambda) = \sum_{{\bf a} \setminus a_r} \pi_{{\bf a}},~\forall~r,a_r$.
\end{definition}
\noindent Here, ${\bf a} := [a_1, \dots, a_\mathrm{N}] \in \{+1,-1\}^\mathrm{N}$, where $\mathrm{N}$ is the cardinality of the set $S$, and ${\bf a} \setminus a_r$ denotes the summation over all components except $a_r$. While sharp spin observables along distinct directions are not jointly measurable, their unsharp versions can be, for suitable values of $\lambda$. For instance, spin observables along the $x$ and $y$ directions are jointly measurable when $\lambda \leq 1/\sqrt{2}$, and observables along $x$, $y$, and $z$ directions are jointly measurable when $\lambda \leq 1/\sqrt{3}$ \cite{Busch1986} (see also \cite{Liang2011, Pal2011, Uola2016}).

\begin{figure}[t!]
\centering
\subfloat[Naive strategy]{\includegraphics[width=0.48\linewidth]{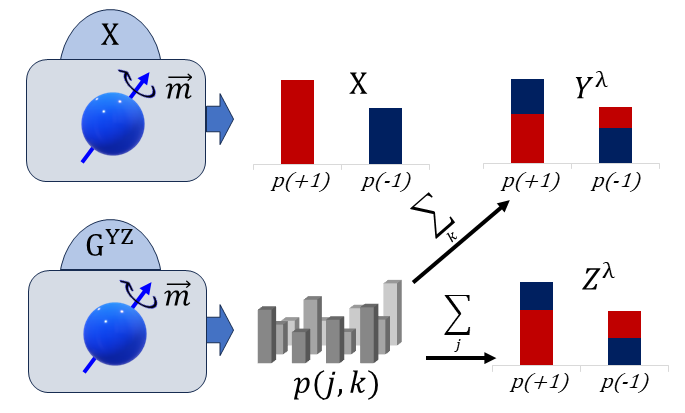}}
\hspace{0.1cm}
\subfloat[Strategy of Carmeli et al.]{\includegraphics[width=0.48\linewidth]{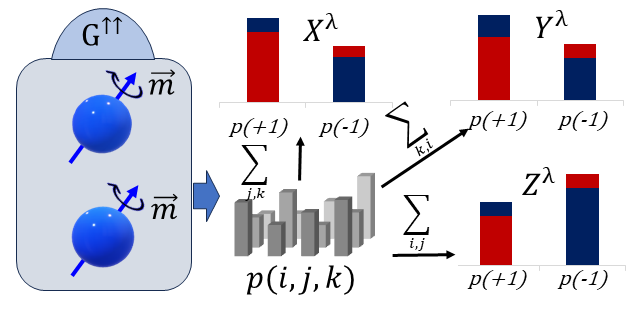}}\\
\vspace{-0.3cm}
\subfloat[Antiparallel configuration]{\includegraphics[width=1\linewidth]{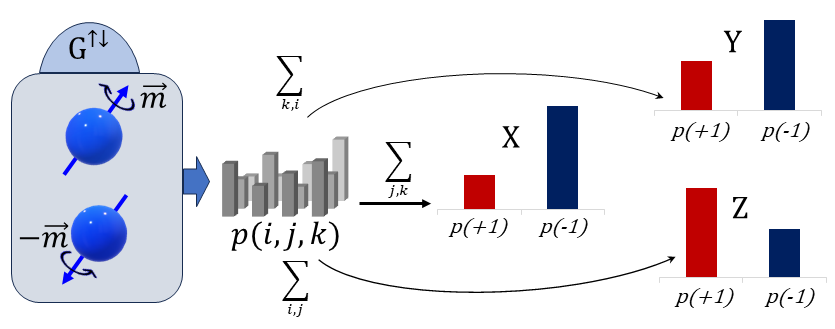}}
\caption{(Color online) (a) Naive strategy: Observable \(X\) is measured on one copy with red and blue bars respectively denoting the probability of outcomes \(+1~\&~-1\) on the given state. The other two observables are jointly measured on the remaining copy with sharpness value \(\lambda=1/\sqrt{2}\) \cite{Busch1986}, with blue (red) part in the red (blue) bar indicating the `unsharpness' in the corresponding outcome. (b) Parallel configuration: All three observables can be jointly measured on two-copies with sharpness value \(\lambda=\sqrt{3}/2\) \cite{Carmeli2016} (b) Antiparallel configuration: All three observables can be jointly measured with sharpness value \(\lambda=1\) [Theorem \ref{theo1}] }\label{Fig1}
\vspace{-.35cm}
\end{figure}

{\it Measurement compatibility in multi-copy setting.--} More recently, Carmeli et al. \cite{Carmeli2016} have investigated measurement compatibility in a multi-copy setting, where the experimenter has access to multiple copies of a quantum state in each measurement run. Notably, any set of $N$ spin observables is $\mathrm{K}$-copy jointly measurable whenever $\mathrm{K} \ge N$, by simply performing different measurements on different copies of the state. However, the scenario becomes nontrivial when $\mathrm{K} < N$. As shown in \cite{Carmeli2016}, the three mutually unbiased spin observables $X, Y, Z$ (corresponding to directions $\hat{x}, \hat{y}, \hat{z}$) are 2-copy jointly measurable for sharpness values up to $\lambda = \sqrt{3}/2$, whereas a naive strategy achieves joint measurability only up to $\lambda = 1/\sqrt{2}$ (see Fig.\ref{Fig1}). We now turn to the anti-parallel ({\footnotesize`\(\uparrow\hspace{-.05cm}\downarrow\)`}) configuration, where the measurement is performed on $\rho_{\vec{m}} \otimes\mathrm{F}(\rho_{\vec{m}})\equiv\rho_{\vec{m}} \otimes \rho_{-\vec{m}}$, with $\rho_{-\vec{m}}:= \tfrac{1}{2}(\id - \vec{m} \cdot \vec{\sigma})$. This leads us to the following definition:
\begin{definition}\label{def3}
A set $S$ of $N$ spin observables is jointly measurable on 2-copy {\footnotesize$\uparrow\hspace{-.05cm}\downarrow$} configuration if there exists a POVM $\mathcal{G}^{_{\uparrow\hspace{-.05cm}\downarrow}} \equiv \{\pi^{_{\uparrow\hspace{-.05cm}\downarrow}}_{{\bf a}} \in \mathcal{P}((\mathbb{C}^2)^{\otimes 2})~|~ \sum_{{\bf a}} \pi^{_{\uparrow\hspace{-.05cm}\downarrow}}_{{\bf a}} = \id^{\otimes 2}\}$ such that, for all states $\rho_{\vec{m}}$ and for all $r$, $\mathrm{Tr}[\rho_{\vec{m}}\, \mathrm{P}^{a_r}_{\hat{n}_r}(\lambda)] = \sum_{{\bf a} \setminus a_r} \mathrm{Tr}[(\rho_{\vec{m}} \otimes \rho_{-\vec{m}})\, \pi^{_{\uparrow\hspace{-.05cm}\downarrow}}_{{\bf a}}]$.
\end{definition}
\noindent In what follows, we present a necessary and sufficient condition for joint measurability of spin observables in the {\footnotesize$\uparrow\hspace{-.05cm}\downarrow$}-configuration.

\begin{proposition}\label{prop1}
A set of spin observables $S = \{\sigma_{\hat{n}_r}(\lambda)\}_r$ is jointly measurable on 2-copy {\footnotesize$\uparrow\hspace{-.05cm}\downarrow$} configuration if and only if there exists a POVM $\mathcal{G}^{_{\uparrow\hspace{-.05cm}\downarrow}} \equiv \{\pi^{_{\uparrow\hspace{-.05cm}\downarrow}}_{{\bf a}} ~|~\sum_{{\bf a}} \pi^{_{\uparrow\hspace{-.05cm}\downarrow}}_{{\bf a}} = \id^{\otimes 2}\}$, such that $\mathbb{P}^{a_r}_{\hat{n}_r}(\lambda) = \sum_{{\bf a} \setminus a_r} \pi^{_{\uparrow\hspace{-.05cm}\downarrow}}_{{\bf a}},~ \forall~r,~a_r$; where $\mathbb{P}^{a_r}_{\hat{n}_r}(\lambda) := \tfrac{1}{2} ( \mathrm{P}^{a_r}_{\hat{n}_r}(\lambda) \otimes \id + \id \otimes \mathrm{F}[\mathrm{P}^{a_r}_{\hat{n}_r}(\lambda)] )$.
\end{proposition}
\noindent Deferring the proof of Proposition \ref{prop1} to Supplemental material \cite{Supple}, here we demonstrate joint measurability of two important sets of spin observables.

\begin{theorem}\label{theo1}
The set $\mathrm{MUB} \equiv \{X, Y, Z\}$ of three spin observables is jointly measurable on 2-copy {\footnotesize $\uparrow\hspace{-.05cm}\downarrow$} configuration.
\end{theorem}
\begin{proof}
(Outline) A straightforward calculation confirms the set of operators 
\begin{align}
\Pi^{_{\uparrow\hspace{-.05cm}\downarrow}}_{[a,b,c]}&:= \tfrac{1}{16}\Bigl(2\id^{\otimes 2} + a\dcb{X,\id}~+ b\dcb{Y,\id} ~+ c\dcb{Z,\id}~ \nonumber\\
& - ab\dacb{X,Y}~ - bc\dacb{Y,Z} ~- ca\dacb{Z,X}~ \Bigr), \label{anti}
\end{align}
to be positive for $~a,b,c = \pm1$; here $\dcb{U,V}~:= U \otimes V - V \otimes U$ and $\dacb{U,V}~:= U \otimes V + V \otimes U$. Furthermore, $\sum_{a,b,c} \Pi_{[a,b,c]}^{_{\uparrow\hspace{-.05cm}\downarrow}} = \id^{\otimes 2}$, ensure $\mathcal{G}^{\uparrow\hspace{-.05cm}\downarrow}_{\mathrm{MUB}} \equiv \big\{\Pi_{[a,b,c]}^{_{\uparrow\hspace{-.05cm}\downarrow}}~|~a,b,c=\pm1\big\}$ to be a POVM on $\mathbb{C}^2\otimes\mathbb{C}^2$. Summing over $b$ and $c$ yields $\sum_{b,c} \Pi_{[a,b,c]}^{_{\uparrow\hspace{-.05cm}\downarrow}} = \tfrac{1}{4} \left( 2 \id^{\otimes 2} + a\, \dcb{X,\id} \right) = \mathbb{P}^a_{\hat{x}}$. Similarly, summing over $(c,a)$ and $(a,b)$ yields $\mathbb{P}^b_{\hat{y}}$ and $\mathbb{P}^c_{\hat{z}}$, respectively. Thus, $\mathcal{G}^{\uparrow\hspace{-.05cm}\downarrow}_{\mathrm{MUB}}$ satisfies the joint measurability condition of Proposition \ref{prop1}, thereby completing the proof.
\end{proof}
\noindent As discussed in the End Matter, symmetries play a crucial role in the explicit construction of the joint POVM in Theorem~\ref{theo1}. In the Supplemental Material~\cite{Supple}, we generalize this construction to a class of three symmetric spin observables. In the next, we rather present joint measurability of four symmetric spin observables on antiparallel configuration.
\begin{theorem}\label{theo2}
The set $\mathrm{SIC}_4 \equiv \{\sigma_{\hat{n}_r}\}_{r=0}^3$ of four qubit observables, along the directions {\footnotesize$\hat{n}_r := \frac{(-1)^{\bar{\delta}_{r,0}}}{\sqrt{3}}\big((-1)^{\delta_{r,1}}, (-1)^{\delta_{r,2}}, (-1)^{\delta_{r,3}}\big)$}, is jointly measurable on 2-copy {\footnotesize $\uparrow\hspace{-.05cm}\downarrow$} configuration.
\end{theorem}
\begin{proof}
(Outline) The set of operators $\mathcal{G}^{\uparrow\hspace{-.05cm}\downarrow}_{\mathrm{SIC}_4} \equiv\Big\{\Pi^{_{\uparrow\hspace{-.05cm}\downarrow}}_{[\Gamma,a]}~|~a = \pm1,\ \Gamma\in \{X,Y,Z\}\Big\}$, where
{\footnotesize\begin{align}
\Pi^{_{\uparrow\hspace{-.05cm}\downarrow}}_{[\Gamma,a]}:= \frac{1}{12} \Bigl(2 \id^{\otimes 2} -2\Gamma^{\otimes2}
+a\sqrt{3}\dcb{\Gamma,\id}~ + \sum_{\Omega\in\{\Gamma\}^{\footnotesize\complement}}\Omega^{\otimes2}\Bigr), \label{sic}
\end{align}}
forms a POVM on $\mathbb{C}^2\otimes\mathbb{C}^2$; {\footnotesize$\complement$} denotes set complement. Furthermore, $\sum_{\Gamma} \Pi^{_{\uparrow\hspace{-.05cm}\downarrow}}_{[\Gamma,+1]} =\tfrac{1}{2}\Bigl(\id^{\otimes2}+\tfrac{1}{2\sqrt{3}}\sum_\Gamma\dcb{\Gamma,\id}~\Bigr)=\mathbb{P}^{+1}_{\hat{n}_0}$. Similarly, summing over the suitable indices reconstructs all $\mathbb{P}^{a_r}_{\hat{n}_r}$ for $r=0,\cdots3,~\&~a_r=\pm1$, satisfying the condition of Proposition \ref{prop1}; and hence completes the proof.
\end{proof}

\noindent We recall here the result of Gisin and Popescu \cite{Gisin1999} which reports that the {\footnotesize$\uparrow\hspace{-.05cm}\downarrow$} spin configuration encodes more classical information than the {\footnotesize$\uparrow\hspace{-.05cm}\uparrow$} configuration. Theorems~\ref{theo1} and~\ref{theo2} show that this advantage extends to the joint measurability of multiple spin observables. Using semidefinite programming, we further establish that this advantage is generic for any triple of spin observables, except when they lie in a common plane (see End Matter).

The spin-flip map $\mathrm{F}$ appearing in the {\footnotesize$\uparrow\hspace{-.05cm}\downarrow$} configuration is a positive trace-preserving (PTP) but not completely positive (CPTP) map \cite{Self6}, which we refer to as PnCP. Moreover, it is self-dual, $\mathrm{F}=\mathrm{F}^{\star}$ \cite{Self4}, and involutive, $\mathrm{F}\circ\mathrm{F}=\mathrm{id}_2$. These properties motivate us to go beyond the specific {\footnotesize$\uparrow\hspace{-.05cm}\uparrow$} and {\footnotesize$\uparrow\hspace{-.05cm}\downarrow$} configurations and study joint measurability for bipartite states of the form $\rho_{\vec m}\otimes\Lambda(\rho_{\vec m})$, where $\Lambda$ is either CPTP or PTP. Although joint measurability with PnCP maps may appear ill-posed, it is operationally meaningful in an \emph{adversarial} setting: a Referee, knowing the state, prepares the configuration and hands it to an experimentalist who is unaware of the preparation but attempts a joint measurement. As we will discuss later, such configurations arise naturally in several practical settings. We first establish a general result on joint measurability in this broader framework; its proof is deferred to the End Matter.
\begin{theorem}\label{theo3}
For any set of spin observables, the joint measurability relation on the configuration $\rho_{\vec{m}} \otimes \Lambda(\rho_{\vec{m}})$ coincides with that of the {\footnotesize$\uparrow\hspace{-.05cm}\uparrow$} configuration whenever $\Lambda$ is a CPTP map.
\end{theorem}
\noindent An advantage in joint measurability over the {\footnotesize$\uparrow\hspace{-.05cm}\uparrow$} configuration, thus, arises only for states of the form $\rho_{\vec{m}} \otimes \Lambda(\rho_{\vec{m}})$ with $\Lambda$ PnCP. A representative family is $\mathrm{F}_\mu(\rho_{\vec{m}})=\tfrac12(\id-\mu\vec{m}\cdot\vec{\sigma})=\tfrac{1-\mu}{2}\id_2+\tfrac{1+\mu}{2}\mathrm{F},\quad \mu\in[-1,1]$, which is PnCP for $\mu\in(1/3,1]$ and CPTP otherwise \cite{Buzek1999,DeMartini2002,Ricci2004}.
\begin{corollary}\label{coro1}
The configuration $\rho_{\vec{m}}\otimes\mathrm{F}_\mu(\rho_{\vec{m}})$ outperforms the {\footnotesize$\uparrow\hspace{-.05cm}\uparrow$} configuration in the joint measurability of unsharp MUB observables whenever $\mu>\sqrt3-1$.
\end{corollary}
\noindent The proof is deferred to the End Matter. We emphasize, however, that PnCP maps do not generically guarantee an advantage as in Corollary~\ref{coro1}. For example, consider the PnCP pancake map $\Lambda^{\mathrm{PC}}(\rho_{\vec{m}}):=\tfrac12(\rho_{\vec{m}}+\rho_{\vec{m}}^{\mathrm T})=\tfrac12(\id+m_x X+m_z Z)$, where $\mathrm{T}$ denotes transposition in the computational basis. \begin{observation}
The spin observables $Y$ and $(Z+Y)/\sqrt2$ are jointly measurable on the {\footnotesize$\uparrow\hspace{-.05cm}\uparrow$} configuration, but not on $\rho_{\vec{m}}\otimes\Lambda^{\mathrm{PC}}(\rho_{\vec{m}})$.
\end{observation}
\noindent Joint measurability on the {\footnotesize$\uparrow\hspace{-.05cm}\uparrow$} configuration is trivial. In contrast, for the eigenstates of $Y\equiv \rho_Y^\pm=\tfrac12(\id\pm Y)$ one has $\rho_Y^\pm\otimes\Lambda^{\mathrm{PC}}(\rho_Y^\pm)=\tfrac12(\id\pm Y)\otimes\tfrac{\id}{2}$, which shows that a sharp $Y$ statistics can be obtained only if $Y$ is performed on the first copy, while the second copy lacks sharp statistics for $(Z+Y)/\sqrt2$. Thus whether $\mathrm{F}_\mu$ yields an advantage in the intermediate regime $\mu\in(1/3,\sqrt{3}-1]$ for other choices of observables remains an open question.

The enhanced measurement compatibility in the {\footnotesize$\uparrow\hspace{-.05cm}\downarrow$} configuration (and in Corollary~\ref{coro1}) can be better appreciated within the framework of generalized probabilistic theories (GPTs) \cite{Ludwig1967,Ludwig1968,Mielnik1968,Mackey1977,Hardy2001,Barrett2007,Chiribella2011,Hardy2011}. Assuming local systems are quantum, GPTs allow multiple prescriptions for composing subsystems, interpolating between minimal and maximal tensor products \cite{Namioka1969,Klay1987,Barnum2010,Torre2012,Naik2022,Patra2023}. The minimal tensor product—containing only separable states but an enlarged set of effects—permits composite measurements forbidden in standard quantum theory, rendering certain incompatible observables jointly measurable. In particular, we establish the following result.
\begin{theorem}\label{theo4}
For any set of qubit POVMs, joint measurability on the {\footnotesize$\uparrow\hspace{-.05cm}\downarrow$} configuration in quantum theory implies joint measurability on the {\footnotesize$\uparrow\hspace{-.05cm}\uparrow$} configuration in minimal tensor product composition. 
\end{theorem}
\noindent The proof is deferred to the Supplemental Material, and there we also note that the converse statement of Theorem \ref{theo4} does not hold in general. In the End Matter, we introduce joint measurability restricted to a sub-ensemble $\mathcal{D}_{\mathrm{SE}}\subset\mathcal{D}(\mathbb{C}^2)$, providing an experimentally accessible route to test the advantages established in Theorems~\ref{theo1}, \ref{theo2}, and Corollary~\ref{coro1}. The remainder of the manuscript shows connections to foundational concepts and potential applications.

{\it Mean King problem and Bub's Cryptography.--} The mean King problem, proposed by Vaidman, Aharonov, and Albert (VAA) \cite{Vaidman1987} (see also \cite{Aharonov2001}) involves a retrodiction task, in which Alice prepares a spin-½ system and sends it to Bob (the `mean King'), who measures one of the observables $X$, $Y$, or $Z$ without revealing his choice. Upon receiving back the particle from Bob, Alice performs a second measurement and correctly retrodicts what Bob’s result would have been for each possible measurement choice. While trivial for two observables—by preparing an eigenstate of one and measuring along the other—the case of three observables is nontrivial. VAA introduced a clever strategy to succeed in this setting. Alice prepares the singlet state $\ket{\psi^-} := \tfrac{1}{\sqrt{2}}(\ket{0}_A\ket{1}_C - \ket{1}_A\ket{0}_C)$ \cite{Self5}, sending the channel particle (C) to Bob and retaining the ancilla (A). Bob measures $X$, $Y$, or $Z$ on particle C and returns it to Alice.

Depending on Bob's measurement, the particles collapse to $\ket{\eta}_A\ket{\eta^\perp}_C$, where $\ket{\eta^\perp}$ is the eigenstate corresponding to Bob measurement outcome. On the combined system Alice performs a measurement $\mathbb{M}_{+}:= \{\xi_{[a,b,c]} ~|~ abc = + 1\}$, where $\xi_{[a,b,c]} = \ket{\xi}_{[a,b,c]}\bra{\xi}\propto\Pi^{_{\uparrow\hspace{-.05cm}\downarrow}}_{[a,b,c]}$ with  $\ket{\xi}_{[a,b,c]} := \tfrac{1}{2}(\ket{\psi^-} - a\ket{\phi^-} + \textbf{i}b\ket{\phi^+} + c\ket{\psi^+})$. Here $\{\ket{\phi^\pm}, \ket{\psi^\pm}\}$ are the Bell states. Based on the outcome, she obtains a look-up table (see Table \ref{tab1})  and successfully retrodicts Bob’s result. Notably, the measurement $\mathcal{G}^{_{\uparrow\hspace{-.05cm}\downarrow}}_{\mathrm{MUB}}$ in Theorem \ref{theo1} can be thought as probabilistic mixture of two projective measurements: $\mathcal{G}^{_{\uparrow\hspace{-.05cm}\downarrow}}_{\mathrm{MUB}} = \tfrac{1}{2} \mathbb{M}_{+} + \tfrac{1}{2} \mathbb{M}_{-}$, where $\mathbb{M}_{\pm}:= \{\xi_{[a,b,c]} ~|~ abc = \pm 1\}$. Likewise $\mathbb{M}_{+}$, the measurement $\mathbb{M}_{-}$ also perfectly wins `mean-King' challenge, thereby ensuring that the POVM $\mathcal{G}^{_{\uparrow\hspace{-.05cm}\downarrow}}_{\mathrm{MUB}}$ also do the same (see Table \ref{tab1}). At this point, we also note that the measurement $\mathbb{M}_{+}$ has also been studied from a different angle by Gisin, who called it the `Elegant Joint Measurement' (EJM) \cite{Gisin2019}. In Supplemental \cite{Supple} we further show that the POVM $\mathcal{G}^{_{\uparrow\hspace{-.05cm}\downarrow}}_{\mathrm{SIC}_4}$ (appears in Theorem~\ref{theo2}) allows Alice to win the mean-King retrodiction task when his measurements are chosen from the set $\mathrm{SIC}_4$. 

Subsequent to VAA's work, Mermin has pointed out that the provocative title of VAA's paper suggests—without asserting—that Alice determines simultaneous values of three orthogonal spin components, although she does not \cite{Mermin1995}. Her measurement yields three candidate outcomes, only one of which—the one corresponding to Bob’s actual choice—matches his result; the others do not correspond to any measurement actually performed (thus invoking an element of counter-factuality). In contrast, as established in our Theorem \ref{theo1}, the POVM $\mathcal{G}^{_{\uparrow\hspace{-.05cm}\downarrow}}_{\mathrm{MUB}}$ implements a joint measurement of all three spin observables on two particles prepared in the {\footnotesize$ \uparrow\hspace{-.05cm}\downarrow$} configuration.
\begin{table}[t!]
\begin{tabular}{|c||c|c|c|c|c|}
\hline
 &  & Outcome $\backslash$ Retrodiction & \cellcolor{blue!35}~~~~$X$~~~~ & ~~~~$Y$~~~~ & \cellcolor{blue!35}~~~~$Z$~~~~ \\ \hline\hline
\multirow{8}{*}{~$\mathcal{G}^{\uparrow\hspace{-.05cm}\downarrow}_{\mathrm{MUB}}$~} & \multirow{4}{*}{$\mathbb{M}_{+}$} & \cellcolor{yellow!50}$\xi_{[+1,+1,+1]}$ & \cellcolor{green!50}$-1$ & $-1$ & \cellcolor{green!50}$-1$ \\ \cline{3-6} 
&  & \cellcolor{purple!15}$\xi_{[+1,-1,-1]}$ & \cellcolor{red!50}$-1$ & $+1$ & \cellcolor{red!50}$+1$ \\ \cline{3-6} 
&  & \cellcolor{purple!15}$\xi_{[-1,+1,-1]}$ & \cellcolor{green!50}$+1$ & $-1$ & \cellcolor{green!50}$+1$ \\ \cline{3-6} 
&  & \cellcolor{purple!15}$\xi_{[-1,-1,+1]}$ & \cellcolor{red!50}$+1$ & $+1$ & \cellcolor{red!50}$-1$ \\ \cline{2-6}\cline{2-6}\cline{2-6}\cline{2-6}\cline{2-6} 
& \multirow{4}{*}{$\mathbb{M}_{-}$} & \cellcolor{purple!15}$\xi_{[-1,+1,+1]}$ & $+1$ & $-1$ & $-1$ \\ \cline{3-6} 
&  & \cellcolor{purple!15}$\xi_{[+1,-1,+1]}$ & $-1$ & $+1$ & $-1$ \\ \cline{3-6} 
&  & \cellcolor{purple!15}$\xi_{[+1,+1,-1]}$ & $-1$ & $-1$ & $+1$ \\ \cline{3-6} 
&  & \cellcolor{yellow!50}$\xi_{[-1,-1,-1]}$ & $+1$ & $+1$ & $+1$ \\ \hline
\end{tabular}
\caption{(Color online) To win the mean King’s challenge, Alice performs the measurement $\mathrm{M}_+ = \{\xi_{[a,b,c]}~|~abc = +1\}$ on her ancilla and the channel system received from Bob. Upon obtaining the outcome $\xi_{[a,b,c]}$, she retrodicts Bob’s outcomes for $X$, $Y$, and $Z$ as $-a$, $-b$, and $-c$, respectively. When Bob restricts his measurements to $X$ and $Z$, a secure key can be established: outcomes corresponding to $\xi_{[+1,-1,-1]}$ and $\xi_{[-1,-1,+1]}$ (marked in red) are used for eavesdropping detection, while the correlation in the remaining outcomes (marked in green) generates the raw key. Although such perfect correlations are absent in $\mathrm{M}_+$ when all three observables are included, a key can still be distilled from the outcomes $\xi_{[+1,+1,+1]}$ and $\xi_{[-1,-1,-1]}$ of the POVM $\mathcal{G}^{_{\uparrow\hspace{-.05cm}\downarrow}}_{\mathrm{MUB}}$ (highlighted in yellow), with the other outcomes (highlighted in purple) serving to detect potential interference by Eve.}\label{tab1}
\vspace{-.35cm}
\end{table}

Building on the mean King strategy of VAA, Bub proposed a quantum key distribution (QKD) protocol in which Bob restricts his measurements to two observables (e.g., \(X\) and \(Z\)), while Alice performs the POVM \(\mathbb{M}_{+}\) jointly on her ancilla and the particle returned by Bob \cite{Bub2001} (see also \cite{Werner2009}). To detect eavesdropping, Alice publicly announces her retrodicted outcomes whenever the projectors \(\xi_{[+1,-1,-1]}\) or \(\xi_{[-1,-1,+1]}\) occur. Bob then verifies whether her retrodiction agrees with his actual measurement outcome. In the absence of discrepancies---indicating no intervention by an eavesdropper---the remaining outcomes are used to generate a shared raw key (see Table~\ref{tab1}). Unlike the BB84 \cite{Bennett2014} and E91 \cite{Ekert1991} protocols, this scheme enables key generation without revealing measurement bases, provided Bob’s choices are restricted to two observables. If Bob instead chooses among all three Pauli observables \(X, Y,\) and \(Z\), basis disclosure is generally required. Remarkably, as shown in Table~\ref{tab1}, the POVM \(\mathcal{G}^{_{\uparrow\hspace{-.05cm}\downarrow}}_{\mathrm{MUB}}\) enables secure key generation even in this three-observable setting.

We now consider the case where Bob performs unsharp spin measurements. As shown in \cite{Busch1986}, the pair \(\{X(\lambda), Z(\lambda)\}\) is jointly measurable for \(\lambda \le 1/\sqrt{2}\), whereas the triple \(\{X(\lambda), Y(\lambda), Z(\lambda)\}\) is jointly measurable only for \(\lambda \le 1/\sqrt{3}\). Consequently, for \(\lambda \in [0,1/\sqrt{2}]\), the original two-observable protocol lies entirely within a compatibility regime and therefore yields no key. In contrast, the modified scheme employing three unsharp observables accesses the intermediate regime \(\lambda \in (1/\sqrt{3}, 1/\sqrt{2}]\), where pairwise incompatibility persists while triple-wise compatibility is absent. This identifies a parameter range in which the enhanced compatibility structure enabled by the {\footnotesize$\uparrow\hspace{-.05cm}\downarrow$} configuration promises improved noise robustness to implement Bub's protocol. A detailed analysis of key-rate calculations and noise-tolerance bounds is beyond the scope of the present Letter and is left for future work. We rather turn to another potential application of enhanced measurement compatibility in the {\footnotesize$\uparrow\hspace{-.05cm}\downarrow$} configuration.

{\it Estimating unknown measurement devices.--} Estimation of unknown quantum resources—such as states, channels, or measurement devices—is a central task for the reliable implementation of quantum protocols \cite{Poyatos1997,Altepeter2003,Mohseni2008,PARIS2009,Giovannetti2011,Eisert2020}. When a resource is supplied by an untrusted party, a verifier seeks to identify it as efficiently as possible using minimal experimental resources. Here, we consider the task of estimating an unknown qubit projective measurement \(\mathrm{M}_{\hat{n}} \equiv \{\mathrm{P}^{a}_{\hat{n}} \mid a=\pm 1\}\), where the measurement axis \(\hat{n}\) is uniformly distributed over the Bloch sphere. The verifier treats the device as a black box with trusted outcome labels and aims to estimate \(\hat{n}\) using as few invocations of the device as possible.

\noindent In a standard approach, the verifier repeatedly measures a fixed probe state (e.g., \(\ket{0}_P\)). Upon obtaining outcome \(a\), the probe collapses to the corresponding eigenstate \(\ket{\psi_{a\hat{n}}}\) of \(\mathrm{P}^a_{\hat{n}}\). Estimating \(\hat{n}\) then reduces to tomography of the post-measurement state by estimating the expectation values of \(X\), \(Y\), and \(Z\). Concretely, if the black-box measurement is applied to \(3N\) copies of the probe state \(\ket{0}_P\), and the resulting post-measurement states are divided into three ensembles of size \(N\), the expectation values of \(X\), \(Y\), and \(Z\) can be estimated with a reliability \(\mathcal{R}(N)\).

\noindent Alternatively, the verifier may prepare \(N\) copies of the singlet state \(\ket{\psi^-}=\tfrac{1}{\sqrt{2}}(\ket{0}_A\ket{1}_P-\ket{1}_A\ket{0}_P)\) and apply the black-box measurement only to the probe qubit \(P\), leaving the ancilla \(A\) untouched. Conditioned on obtaining outcome \(a=\pm 1\), the post-measurement state becomes \(\ket{\psi_{\mp\hat{n}}}_A \ket{\psi_{\pm\hat{n}}}_P\). Performing the joint POVM \(\mathcal{G}^{_{\uparrow\hspace{-.05cm}\downarrow}}_{\mathrm{MUB}}\) (Theorem~\ref{theo1}) on the bipartite state enables the simultaneous estimation of the expectation values of all three Pauli observables via classical post-processing of the outcome statistics, achieving the same reliability \(\mathcal{R}(N)\). This strategy is resource-efficient in the sense that the black-box measurement device is used only \(N\) times, compared to \(3N\) uses in the unentangled strategy. This indicates a further operational implication of enhanced measurement compatibility in the {\footnotesize$\uparrow\hspace{-.05cm}\downarrow$} configuration. A detailed quantitative analysis, including optimal estimators and experimental implementations—particularly for general unsharp measurement devices—lies beyond the scope of the present Letter and will be addressed in a subsequent work.

{\it Discussions.--} We have explored novel aspects of quantum incompatibility in multi-copy scenarios, demonstrating that the manifestation of measurement incompatibility depends crucially on whether the system copies are prepared in parallel or antiparallel configurations. In particular, we have shown that three mutually orthogonal spin observables -- known to be incompatible in the single -- copy setting—can become jointly measurable when probed on antiparallel spin pairs, which highlights a distinctive feature of the antiparallel configuration. We further connected our findings to other foundational concepts and explored their potential applications. Specifically, we established a link with the mean King retrodiction task, and discussed how the observed compatibility structure plays a critical role in a quantum key distribution  protocol inspired from it. Additionally, we discussed how the joint measurability of three orthogonal spin observables in the antiparallel setup can enhance the efficiency of estimating an unknown measurement device.

\noindent Our study welcomes several questions for future investigation. While Theorems \ref{theo1} and \ref{theo2} demonstrate the advantage of the antiparallel configuration for specific choices of spin observables, it remains an open question whether similar advantages persist for more general sets of observables. Furthermore, Corollary \ref{coro1} establishes an advantage of the configuration $\rho_{\vec{m}} \otimes \Lambda(\rho_{\vec{m}})$ over the parallel configuration for a certain class of PnCP maps. It would be interesting to explore for which class of PnCP maps such an advantage will be possible.  Finally, extending the analysis to higher-dimensional quantum systems may reveal additional structures and potential applications.

\begin{acknowledgements}
\noindent{\bf Acknowledgement}: We gratefully acknowledge insightful discussions with Nicolas Gisin on connection between the POVM $\mathcal{G}^{_{\uparrow\hspace{-.05cm}\downarrow}}_{\mathrm{MUB}}$ and the elegant joint measurement. KA acknowledges support from the CSIR project $09/0575(19300)/2024$-EMR-I. SRC acknowledges support from University Grants Commission, India (Reference no. 211610113404). SGN acknowledges support from the CSIR project $09/0575(15951)/2022$-EMR-I. MB acknowledge the financial support through the National Quantum Mission (NQM) of the Department of Science and Technology, Government of India.
\end{acknowledgements}

%

\newpage
\noindent {\large\bf End Matter}\\\\
\noindent{\bf Symmetry of POVM in Theorem \ref{theo1}:}\\
\noindent The proof of Theorem~\ref{theo1} is constructive and exploits the symmetry of the Pauli mutually unbiased bases ${X,Y,Z}$:
\begin{itemize}[itemsep=2pt,leftmargin=1.5em]    \item Firstly, Joint measurability implies that each POVM element $\Pi^{_{\uparrow\hspace{-.05cm}\downarrow}}_{[a,b,c]}$ is of rank at most one; e.g., $\Pi^{_{\uparrow\hspace{-.05cm}\downarrow}}_{[+1,+1,+1]}$ has no support on the $-1$ eigen-spaces of $X,Y,Z$ in the antiparallel configuration.
\item The POVM can be chosen to satisfy outcome-flip symmetry, $\Pi^{_{\uparrow\hspace{-.05cm}\downarrow}}_{[-a,-b,-c]}=\mathrm{F}\otimes\mathrm{F}\big(\Pi^{_{\uparrow\hspace{-.05cm}\downarrow}}_{[a,b,c]}\big)$, where $\mathrm{F}$ denotes the spin-flip operation.
\item In addition, invariance under permutations of the Pauli axes implies $\Pi^{_{\uparrow\hspace{-.05cm}\downarrow}}_{[a,b,c]}=\mathrm{R_{XY}}\otimes\mathrm{R_{XY}}\big(\Pi^{_{\uparrow\hspace{-.05cm}\downarrow}}_{[b,a,c]}\big)=\mathrm{R_{YZ}}\otimes\mathrm{R_{YZ}}\big(\Pi^{_{\uparrow\hspace{-.05cm}\downarrow}}_{[a,c,b]}\big)=\mathrm{R_{ZX}}\otimes\mathrm{R_{ZX}}\big(\Pi^{_{\uparrow\hspace{-.05cm}\downarrow}}_{[c,b,a]}\big)$, with $\mathrm{R_{XY}}$ exchanging $X$ and $Y$ and leaving $Z$ invariant, and $\mathrm{R_{YZ}}$, $\mathrm{R_{ZX}}$ defined analogously.
\end{itemize}
These symmetries imply that it suffices to specify only two POVM elements, $\Pi^{_{\uparrow\hspace{-.05cm}\downarrow}}_{[+1,+1,+1]}$ and $\Pi^{_{\uparrow\hspace{-.05cm}\downarrow}}_{[+1,+1,-1]}$, from which all remaining elements are uniquely fixed. A detailed proof of the symmetry properties for a class of symmetric observable sets is provided in Lemma~1S of Supplemental Material.\\
   
\noindent {\bf Proof of Theorem \ref{theo3}:}
\vspace{-.3cm}
\begin{proof}
Suppose that the configuration $\rho_{\vec{m}} \otimes \Lambda(\rho_{\vec{m}})$ admits a POVM $\mathcal{G} \equiv \{\pi_{{\bf a}}~|~ \sum_{{\bf a}} \pi_{{\bf a}} = \id^{\otimes 2}\}$ ensuring joint measurability of the observables $\{\sigma_{\hat{n}_r}(\lambda)\}_r$. Then, $
\mathrm{Tr}[\rho_{\vec{m}}\, \mathrm{P}_{\hat{n}_r}^{a_r}(\lambda_{\text{opt}})] = \sum_{{\bf a} \setminus a_r} \mathrm{Tr}[\rho_{\vec{m}} \otimes \Lambda(\rho_{\vec{m}})\, \pi_{{\bf a}}]$, $\forall~r,~a_r,~\rho_{\vec{m}}$. Using the identity $\mathrm{Tr}[\rho_{\vec{m}} \otimes \Lambda(\rho_{\vec{m}})\, \pi_{{\bf a}}] = \mathrm{Tr}[\rho_{\vec{m}} \otimes \rho_{\vec{m}}\, \{\text{id}_2 \otimes \Lambda^\star(\pi_{{\bf a}})\}]$, where $\Lambda^\star$ is the dual map, we define the new POVM $\mathcal{G}' \equiv \{\text{id}_2 \otimes \Lambda^\star(\pi_{{\bf a}})\}_{{\bf a}}$. Since $\Lambda$ is CPTP, its dual $\Lambda^\star$ is CP and unital (i.e., $\Lambda^\star(\id) = \id$) -- ensures that $\mathcal{G}'$ is a valid POVM, which warrants joint measurability of the same set of observables on the {\footnotesize$\uparrow\hspace{-.05cm}\uparrow$} configuration. This completes the proof.
\end{proof}

\noindent {\bf Proof of Corollary \ref{coro1}:}
\vspace{-.3cm}
\begin{proof}
Consider the POVM $\mathcal{G}^{_{\uparrow\hspace{-.05cm}\downarrow}}_{\mathrm{MUB}}$, applied to the configuration $\rho_{\vec{m}} \otimes \mathrm{F}_\mu(\rho_{\vec{m}})$. The probability of clicking the effect $\Pi^{_{\uparrow\hspace{-.05cm}\downarrow}}_{[a,b,c]}$ is $
p(a,b,c)${\footnotesize$ = \Tr\left[\Pi^{_{\uparrow\hspace{-.05cm}\downarrow}}_{[a,b,c]}\, \rho_{\vec{m}} \otimes \mathrm{F}_\mu(\rho_{\vec{m}})\right]$}. Summation over $a,b$ yields $
\sum_{a,b} p(a,b,c) ${\footnotesize$= \tfrac{1}{2}\left(1 + c~\frac{1 + \mu}{2}m_z\right)$}, which matches the outcome statistics of the measurement $\{\mathrm{P}^c_{\hat{z}}((1 + \mu)/2)\}_c$ on state $\rho_{\vec{m}}$. Analogous calculations for the other pairs of indices show that the configuration simulates unsharp MUB observables with effective sharpness $\lambda = (1 + \mu)/2$. Since the {\footnotesize$\uparrow\hspace{-.05cm}\uparrow$} configuration permits joint measurability of MUB observables only up to $\lambda = \sqrt{3}/2$ \cite{Carmeli2016}, the present configuration provides an advantage whenever $\mu > (\sqrt{3} - 1)$.
\end{proof}

\noindent{\bf Optimal sharpness through SDP:}\\
In this section, we formulate a semidefinite programming (SDP) approach to determine the optimal sharpness parameters for both parallel and antiparallel configurations, in direct analogy with the standard notion of joint measurability. For an arbitrary set of observables $S\equiv\{\sigma_{\hat n_r}\}_r$, the optimal sharpness achievable in the antiparallel configuration, denoted $\lambda^{\uparrow\hspace{-.05cm}\downarrow}_{\mathrm{opt}}$, is given by the following SDP:
\begin{align}
&\hspace{3cm}\text{Maximize}:~\lambda \nonumber\\
&\text{Subject to}:~\pi_{\mathbf{a}} \in \mathcal{E}(\mathbb{C}^2 \otimes \mathbb{C}^2),~
\sum_{\mathbf{a}} \pi_{\mathbf{a}} = \id^{\otimes 2},~
\lambda \geq 0; \nonumber\\
&\frac{1}{2} \left(
\mathrm{P}^{a_r}_{\hat{n}_r}(\lambda) \otimes \id
+ \id \otimes \mathrm{F}\!\left[\mathrm{P}^{a_r}_{\hat{n}_r}(\lambda)\right]
\right)
= \sum_{\mathbf{a} \setminus a_r} \pi_{\mathbf{a}}\forall~ r,~ a_r.
\label{sdp0}
\end{align}
Equation~(\ref{sdp0}) follows directly from the necessary and sufficient condition of Proposition~\ref{prop1}. As shown in Ref.~\cite{Carmeli2016}, replacing the flip map $\mathrm{F}$ in Eq.~(\ref{sdp0}) by the identity map yields the corresponding criterion for joint measurability in the parallel configuration. An analogous SDP therefore provides the optimal sharpness parameter $\lambda^{\uparrow\hspace{-.05cm}\uparrow}_{\mathrm{opt}}$ for the parallel case.

\begin{figure}[t!]
\centering
{\includegraphics[width=1\linewidth]{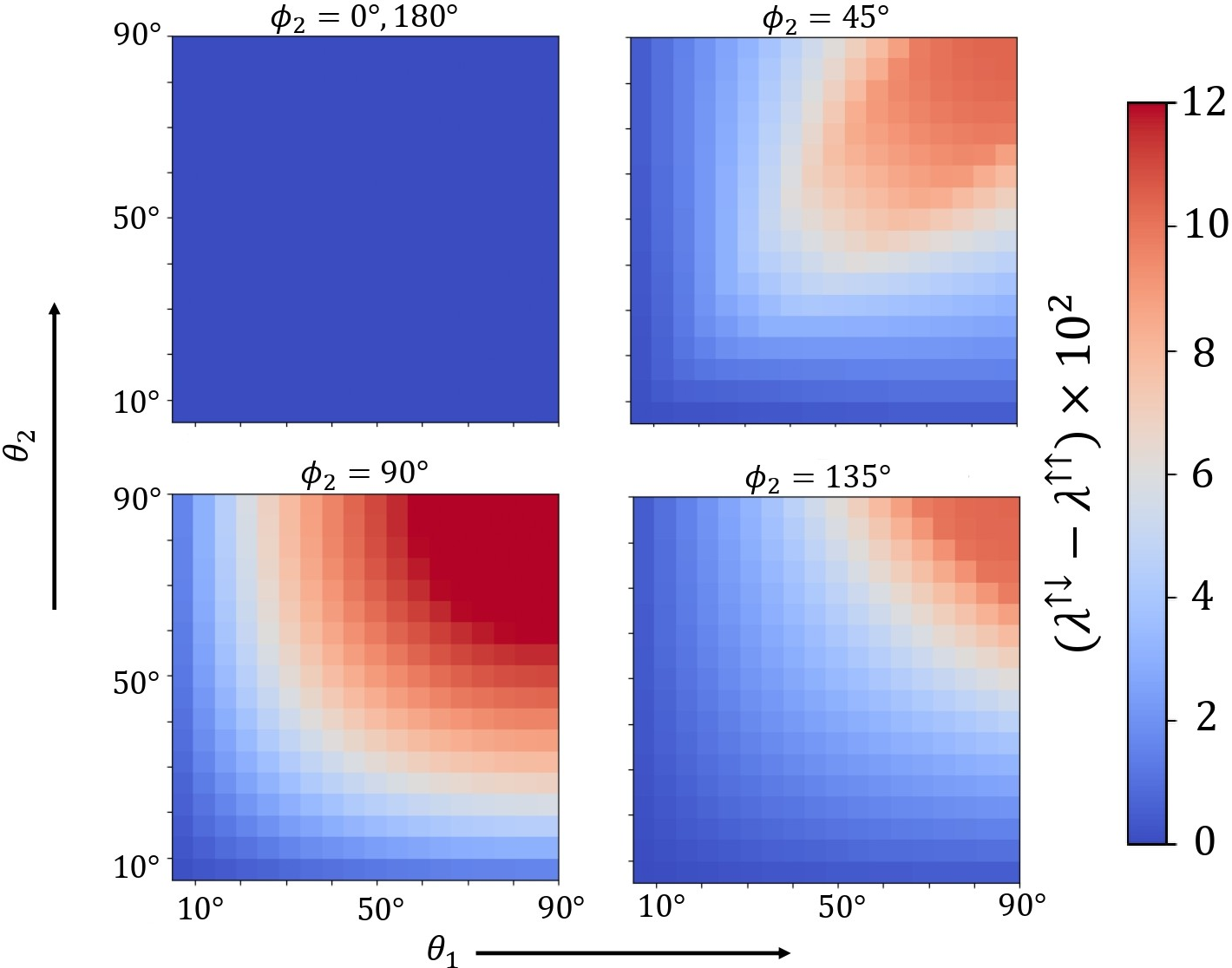}}
\caption{(Color online) The figure shows a heat map of $\lambda^{_{\uparrow\hspace{-.05cm}\downarrow}}_{\mathrm{opt}}-\lambda^{_{\uparrow\hspace{-.05cm}\uparrow}}_{\mathrm{opt}}$ as a function of $\phi_2$. Dark red regions indicate a pronounced advantage, while dark blue regions correspond to negligible advantage. In all cases $\lambda^{_{\uparrow\hspace{-.05cm}\downarrow}}_{\mathrm{opt}} \geq \lambda^{_{\uparrow\hspace{-.05cm}\uparrow}}_{\mathrm{opt}}$, with equality occurring only when the observables lie in a great plane.}\label{heatmap}
\vspace{-.35cm}
\end{figure}

\noindent{\bf Three qubit observables:} Using the SDP developed above, we numerically show that for any triple of spin observables the optimal sharpness parameters satisfy
\begin{align}
\lambda^{\uparrow\hspace{-.05cm}\downarrow}_{\mathrm{opt}} \geq
\lambda^{\uparrow\hspace{-.05cm}\uparrow}_{\mathrm{opt}},  
\end{align}
with equality if and only if the observables lie in a common great plane. We begin, without loss of generality, with the choice of observables $S=\{\sigma_{n_1},\sigma_{n_2},\sigma_{n_3}\}$:
\begin{align}
\left\{\begin{aligned}
\sigma_{n_1}&=Z,\hspace{2cm}
\sigma_{n_2}=\cos\theta_1Z+\sin\theta_1X\nonumber\\
\sigma_{n_3}&=\cos\theta_2Z+\sin\theta_2\cos{\phi_2}X+\sin\theta_2\sin{\phi_2}Y
\end{aligned}\right\}
\end{align}
where, $0\leq\theta_1,\theta_2\leq\pi/2$ and $0\leq\phi_2\leq\pi$. The results are depicted in Fig.~\ref{heatmap}. 

\begin{figure}[t!]
\centering
\subfloat[Ensemble \(\mathcal{D}_{\text{GC}}\)]{\includegraphics[width=0.32\linewidth]{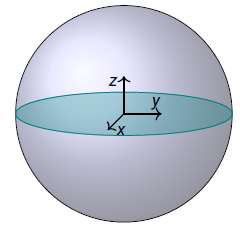}}
\hspace{0.1cm}
\subfloat[Ensemble \(\mathcal{D}_{\text{Tet}}\)]{\includegraphics[width=0.30\linewidth]{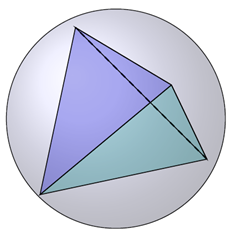}}
\hspace{0.1cm}
\subfloat[Ensemble \(\mathcal{D}_{\text{Oct}}\)]{\includegraphics[width=0.30\linewidth]{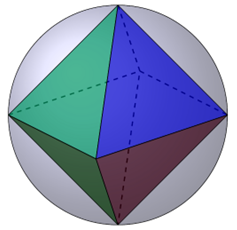}}
\caption{(Color online) On any sub-ensemble \(\mathcal{D}_{\text{GC}}\) on a great circle of the Bloch sphere, and on the sub-ensemble \(\mathcal{D}_{\text{Tet}}\) both the parallel and antiparallel configuration ensure compatibility of \(\{X,Y,Z\}\) for sharpness parameter \(\lambda=1\). On the other hand, on the sub-ensemble \(\mathcal{D}_{\text{Oct}}\) antiparallel configuration ensures compatibility up-to \(\lambda=1\) (a consequence of Theorem \ref{theo1}), whereas parallel configuration allows compatibility up-to \(\lambda\approx0.86603\).\vspace{-.5cm}}\label{Fig2}
\end{figure}

\noindent{\bf Measurement compatibility on sub-ensemble of states:} Joint measurability of a set of spin observables $\mathcal{S}_N\equiv\{P^{a}_{\hat n_r}(\lambda)\}_{r=1}^N$ is conventionally defined via the existence of a parent POVM $\mathcal{G}=\{\pi_{\mathbf a}\}_{\mathbf a}$ reproducing the statistics of each observable for all quantum states through marginalization. This requirement leads to the operator equalities of Definition~\ref{def1}. In many practical situations, however, the input states are drawn from a restricted sub-ensemble $\mathcal{D}_{\mathrm{SE}}\subset\mathcal{D}(\mathbb{C}^2)$, motivating a relaxed notion of joint measurability.
\begin{definition}\label{def4}
A set of spin observables $\mathcal{S}_N$ is jointly measurable on a sub-ensemble $\mathcal{D}_{\mathrm{SE}}\subset\mathcal{D}(\mathbb{C}^2)$ if there exists a POVM $\mathcal{G}=\{\pi_{\mathbf a}\ge0,|,\sum_{\mathbf a}\pi_{\mathbf a}=\id\}$ such that, for all $\rho_{\vec m}\in\mathcal{D}_{\mathrm{SE}}$, $\mathrm{Tr}[\rho_{\vec m}P^{a_r}_{\hat n_r,\lambda}]
=\sum_{\mathbf a\setminus a_r}\mathrm{Tr}[\rho_{\vec m}\pi_{\mathbf a}],~~ \forall,r,a_r$.
\end{definition}
The operator equalities of Definition~\ref{def1} are recovered if and only if the linear span of $\mathcal{D}_{\mathrm{SE}}$ coincides with the space of Hermitian operators on $\mathbb{C}^2$. Consequently, compatibility relations may change when measurements are restricted to a sub-ensemble. For example, the Pauli observables $X,Y,Z$ are incompatible on the full state space, but become jointly measurable on $\mathcal{D}_{\mathrm{SE}}=\mathrm{ConvHull}\{\ket{0}\bra{0},\ket{1}\bra{1}\}$, with the $Z$ measurement acting as a parent POVM and the outcomes of $X$ and $Y$ guessed uniformly at random. By contrast, on the symmetric octahedral ensemble $\mathcal{D}_{\mathrm{Oct}}$ (Fig.~\ref{Fig2}), joint measurability is recovered only up to $\lambda=1/\sqrt3$, matching the unrestricted case. An analogous relaxation applies to Definition~\ref{def3}.

This naturally raises the question of whether there exists a sub-ensemble $\mathcal{D}_{\mathrm{SE}}$ for which the antiparallel configuration offers a genuine advantage over the parallel one in the joint measurability of three mutually orthogonal spin observables. While Theorem~\ref{theo1} guarantees compatibility of $\{X,Y,Z\}$ at $\lambda=1$ in the antiparallel configuration for any sub-ensemble, an experimentally relevant demonstration requires identifying a finite sub-ensemble for which the parallel configuration remains incompatible at $\lambda=1$.

Since compatibility on a sub-ensemble is weaker than on the full state space, the optimal sharpness parameter in the parallel configuration may exceed the unrestricted bound $\lambda=\sqrt3/2$ \cite{Carmeli2016}. For a given set of observables $\mathcal{S}_N$ and a finite sub-ensemble $\mathcal{D}_{\mathrm{SE}}$ of states, the optimal sharpness can be obtained via the SDP
\begin{align}
&\hspace{3cm}\text{maximize } \lambda \nonumber\\
&\text{subject to } \pi_{\mathbf a}\in\mathcal{E}(\mathbb{C}^2\otimes\mathbb{C}^2),~
\sum_{\mathbf a}\pi_{\mathbf a}=\id^{\otimes2},~\lambda\ge0, \nonumber\\
&\mathrm{Tr}[\rho_{\vec m}P^{a_r}_{\hat n_r,\lambda}]
=\sum_{\mathbf a\setminus a_r}\mathrm{Tr}[\rho_{\vec m}^{\otimes2}\pi_{\mathbf a}],
~\forall~r,~a_r,~\rho_{\vec m}\in\mathcal{D}_{\mathrm{SE}} .
\label{sdp}
\end{align}
No advantage is expected for sub-ensembles confined to a great circle, as any such state can be unitarily mapped to its orthogonal counterpart on the same circle \cite{Lo2000,Pati2000,Bennett2001}. A natural discrete and symmetric candidate is the tetrahedral ensemble $\mathcal{D}_{\mathrm{Tet}}:=\{\rho_{\vec m_i}\mid \vec m_i=\tfrac1{\sqrt3}(\pm1,\pm1,\pm1),~m_xm_ym_z=+1\}$. Solving the SDP for $\mathcal{D}_{\mathrm{Tet}}$, we find that the parallel configuration admits a parent POVM $\mathcal{G}_{\mathrm{Tet}}$ that exactly reproduces the statistics of $\{X,Y,Z\}$ at $\lambda=1$.

By contrast, for the octahedral ensemble $\mathcal{D}_{\mathrm{Oct}}$, the SDP yields $\lambda \simeq 0.86603 \simeq \sqrt{3}/2$, in agreement with Ref.~\cite{Carmeli2016}. This identifies $\mathcal{D}_{\mathrm{Oct}}$ as a natural finite sub-ensemble for experimentally demonstrating the enhanced measurement compatibility of the antiparallel configuration. As noted earlier, the POVM in Theorem~\ref{theo1} is a uniform mixture of two mean-King measurements (also called the EJM), $\mathcal{G}^{\uparrow\hspace{-.05cm}\downarrow}_{\mathrm{MUB}} = \tfrac{1}{2}\,\mathbb{M}_{+} + \tfrac{1}{2}\,\mathbb{M}_{-}$, where $\mathbb{M}_{\pm} := \{ \xi_{[a,b,c]} \mid abc = \pm 1 \}$. Since optical implementations of the mean-King task are already available \cite{Oliver2003} and the EJM has recently been implemented in \cite{Huang2022}, these results demonstrate that an experimental verification of the antiparallel advantage in the joint measurability of MUB observables is feasible with current technology.
\newpage
\onecolumngrid
\noindent \begin{center}{\Large \bf Supplemental Material}\end{center}
~\vspace{-0.5cm}
\section{Proof of Proposition 1}
\noindent The notion joint measurability is in general studied for general POVMs. A `$K$' outcome qubit POVM $\mathcal{A}$ is the collection of $K~(<\infty)$ effects adding up to identity operator. Formally
\begin{align}
\mathcal{A}\equiv\left\{\mathcal{A}^r\in\mathcal{E}(\mathbb{C}^2)~|~\sum_{r=1}^K\mathcal{A}^r=\id\right\}.
\end{align}
\begin{manualdefinition}{1S}\label{def1s}
A collection of $N$ qubit POVMs $\{\mathcal{A}_{l}\equiv\{\mathcal{A}_{l}^{r_l}\}_{r_l=1}^{K_l}\}_{l=1}^N$ is said to be jointly measurable on the configuration $\text{id}_2 \otimes \Lambda (\rho_m \otimes \rho_m)$, with  $\Lambda$ being a PTP map, if there exists a POVM $\mathcal{G}\equiv\left\{\Pi_{\bf r}\in\mathcal{E}((\mathbb{C}^2)^{\otimes2})~|~\sum_{\bf r}\Pi_{\bf r}=\id^{\otimes 2}\right\}~\text{with}~{\bf r}:=[r_1,\cdots,r_N]\in K_1\times\cdots\times K_N$, such that for all $\rho_{\vec{m}}\in\mathcal{D}(\mathbb{C}^2)$
\begin{align}
\sum_{{\bf r}\setminus r_l}\Tr\left[\Pi_{\bf r}\left(\rho_{\vec{m}}\otimes \Lambda(\rho_{\vec{m}})\right)\right]=\Tr\left[\mathcal{A}_l^{r_l}\rho_{\vec{m}}\right],~~ \forall~l\in\{1,\cdots,N\}~\&~r_l\in K_l.\label{1reproduceOk}
\end{align}    
\end{manualdefinition}

\noindent The \textbf{Proposition 1} in main manuscript consider only unsharp spin observables and {\footnotesize$\uparrow\hspace{-.05cm}\downarrow$} configuration. Here, we establish the result for more general POVMs.
\begin{manualproposition}{1S}\label{prop1s}
A collection of $N$ qubit POVMs $\{\mathcal{A}_{l}\equiv\{\mathcal{A}_{l}^{r_l}\}_{r_l=1}^{K_l}\}_{l=1}^N$ is jointly measurable on the state configuration $\rho_{\vec{m}}\otimes\mathrm{F}(\rho_{\vec{m}})$ if and only if there exists a POVM $\mathcal{G}\equiv \{\Pi_{{\bf r}}\in\mathcal{E}((\mathbb{C}^2)^{\otimes2}) ~|~\sum_{{\bf r}} \Pi_{{\bf r}} = \id^{\otimes 2}\}$, with ${\bf r}:=[r_1,\cdots,r_N]\in K_1\times\cdots\times K_N$, such that $\mathbb{A}^{r_l}_{l}= \sum_{{\bf r} \setminus r_l} \Pi_{{\bf r}},~ \forall~l,~r_l$; where $\mathbb{A}^{r_l}_{l}:= \frac{1}{2} \left( \mathcal{A}^{r_l}_{l} \otimes \id + \id \otimes \mathrm{F}[\mathcal{A}^{r_l}_{l}] \right)$.
\end{manualproposition}
\begin{proof}
The proof follows a similar structure as of Theorem 1 in Ref.\cite{Carmeli2016}.\\
\noindent\underline{{\bf Sufficiency}}\\
Let there exists a POVM $\mathcal{G}\equiv \{\Pi_{{\bf r}}\in\mathcal{E}((\mathbb{C}^2)^{\otimes2}) ~|~\sum_{{\bf r}} \Pi_{{\bf r}} = \id^{\otimes 2}\}$ such that
\begin{align}
\sum_{{\bf r} \setminus r_l} \Pi_{{\bf r}}=\mathbb{A}^{a_l}_{l}=\frac{1}{2} \left( \mathcal{A}^{r_l}_{l} \otimes \id + \id \otimes \mathrm{F}[\mathcal{A}^{r_l}_{l}] \right),~ \forall~l,~r_l~.    
\end{align}
Since $\text{id}_2$ and $\mathrm{F}$ are self dual, and $\mathrm{F}\circ\mathrm{F}=\text{id}_2$, we thus have
\begin{align}
\tr\Big[\left(\rho_{\vec{m}} \otimes \mathrm{F}(\rho_{\vec{m}})\right)\sum_{{\bf r} \setminus r_l} \Pi_{{\bf r}}\Big]&=\tr\left[\left(\rho_{\vec{m}} \otimes \mathrm{F}(\rho_{\vec{m}})\right)\mathbb{A}_{l}^{r_l}\right]\nonumber\\
&= \frac{1}{2}\tr\left[\left(\rho_{\vec{m}} \otimes \mathrm{F}(\rho_{\vec{m}})\right)\text{id}_2\otimes \mathrm{F}\left(\id\otimes \mathcal{A}_l^{r_l} + \mathcal{A}_l^{r_l} \otimes \id\right)\right]\nonumber \\
&= \frac{1}{2}\tr\left[\left(\rho_{\vec{m}}\otimes  \mathrm{F}\circ\mathrm{F}(\rho_{\vec{m}})\right)\left(\id\otimes \mathcal{A}_l^{r_l} + \mathcal{A}_l^{r_l}\otimes \id\right)\right]\nonumber \\
&= \frac{1}{2}\tr\left[(\rho_{\vec{m}} \otimes \rho_{\vec{m}})\left(\id\otimes \mathcal{A}_l^{r_l} + \mathcal{A}_l^{r_l} \otimes \id\right)\right]= \tr\left[\rho_{\vec{m}}~\mathcal{A}_l^{r_l}\right]. 
\end{align} 
This implies that the POVMs $\{\mathcal{A}_{l}\equiv\{\mathcal{A}_{l}^{r_l}\}_{r_l=1}^{K_l}\}_{l=1}^N$ are jointly measurable on the state configuration $\rho_{\vec{m}}\otimes\mathrm{F}(\rho_{\vec{m}})$.

\noindent\underline{{\bf Necessity}}\\
Let there exists a POVM $\mathcal{G}'\equiv \{\Pi'_{{\bf r}}\in\mathcal{E}((\mathbb{C}^2)^{\otimes2}) ~|~\sum_{{\bf r}} \Pi'_{{\bf r}} = \id^{\otimes 2}\}$ such that
\begin{align}
\tr\Big[\left(\rho_{\vec{m}} \otimes \mathrm{F}(\rho_{\vec{m}})\right)\sum_{{\bf r} \setminus r_l} \Pi'_{{\bf r}}\Big]=\tr\left[\rho_{\vec{m}}~\mathcal{A}_l^{r_l}\right],~\forall~l,~r_l,~\rho_{\vec{m}}.\label{s1} 
\end{align}
$\mathrm{F}$ being a completely co-positive map $\mathrm{F}\otimes\mathrm{F}$ becomes positive \cite{Muller2015}. Furthermore, since composition of two positive maps is positive, thus $(\mathrm{F}\otimes\mathrm{F})\circ\text{Swap}$ is a PTP map. Therefore it implies existence of another POVM 
\begin{subequations}
\begin{align}
\mathcal{G}"\equiv \{(\mathrm{F}\otimes\mathrm{F})\circ\text{Swap}(\Pi'_{{\bf r}})&:=\Pi"_{{\bf r}}\in\mathcal{E}((\mathbb{C}^2)^{\otimes2}) ~|~\sum_{{\bf r}} \Pi"_{{\bf r}} = \id^{\otimes 2}\},~~\text{satisfying}\\
\tr\Big[\left(\rho_{\vec{m}} \otimes \mathrm{F}(\rho_{\vec{m}})\right)\sum_{{\bf r} \setminus r_l} \Pi"_{{\bf r}}\Big]&=\tr\Big[\left(\rho_{\vec{m}} \otimes \mathrm{F}(\rho_{\vec{m}})\right) (\mathrm{F}\otimes\mathrm{F})\circ\text{Swap}\Big(\sum_{{\bf r} \setminus r_l}\Pi'_{{\bf r}}\Big)\Big]\nonumber\\
&=\tr\Big[\left(\mathrm{F}(\rho_{\vec{m}})\otimes\rho_{\vec{m}}\right)\text{Swap}\Big(\sum_{{\bf r} \setminus r_l}\Pi'_{{\bf r}}\Big)\Big]\nonumber\\
&=\tr\Big[\left(\rho_{\vec{m}} \otimes \mathrm{F}(\rho_{\vec{m}})\right)\sum_{{\bf r} \setminus r_l} \Pi'_{{\bf r}}\Big]=\tr\left[\rho_{\vec{m}}~\mathcal{A}_l^{r_l}\right],~\forall~l,~r_l,~\rho_{\vec{m}}.\label{s2} 
\end{align}   
\end{subequations}
Thus $\mathcal{G}"$ also ensures joint measurability of the POVMs $\{\mathcal{A}_l\}_{l=1}^N$. We therefore can construct another POVM $\mathcal{G}$ of the form 
\begin{align}
\mathcal{G}\equiv\Big\{\Pi_{\bf r}:=\frac{1}{2}\Big(\Pi'_{\bf r}+\Pi"_{\bf r}\Big)~|~\Pi_{\bf r}\in\mathcal{E}((\mathbb{C}^2)^{\otimes2})~\&~\sum_{\bf r}\Pi_{\bf r}=\id^{\otimes2}\Big\},~\text{satisfying}\label{s3}\\
\tr\Big[\left(\rho_{\vec{m}} \otimes \mathrm{F}(\rho_{\vec{m}})\right)\sum_{{\bf r} \setminus r_l} \Pi_{{\bf r}}\Big]=\tr\left[\rho_{\vec{m}}~\mathcal{A}_l^{r_l}\right],~\forall~l,~r_l,~\rho_{\vec{m}}.\label{s4} 
\end{align}
Consider now a state $\rho\in\mathcal{D}(\mathbb{C}^2)$ of the form
\begin{align}
\rho=\frac{1}{2}(\id+t~\Delta),\label{s2}
\end{align}
with $\Delta = \Delta^\dagger~\&~\tr[\Delta]=0$ and $|t|\leq 1/\epsilon_{\max}$, where $\epsilon_{\max}$ is the maximum eigenvalue of $\Delta$. Now, using the fact $\mathrm{F}(\Delta)=-\Delta$ and substituting Eq.(\ref{s2}) in Eq.(\ref{s4}) we have
\begin{align}
\tr\Big[\left(\id^{\otimes2}+t(\Delta\otimes\id-\id\otimes\Delta)- t^2\Delta^{\otimes2}\right)\mathbb{A}^{r_l}_l\Big] = 2\Big[\tr[\mathcal{A}_l^{r_l}] + t\tr[\Delta \mathcal{A}_l^{r_l}]\Big],
\end{align}
where $\mathbb{A}^{r_l}_l:=\sum_{{\bf r} \setminus r_l} \Pi_{{\bf r}}$. Comparing the coefficients of the same degree in $t$, we get
\begin{subequations}
\begin{align}
\tr[\mathbb{A}^{r_l}_l]&= 2\tr[\mathcal{A}_l^{r_l}], \label{a1}\\ 
\tr[(\Delta\otimes\id-\id\otimes\Delta)\mathbb{A}^{r_l}_l]&=2\tr[\Delta \mathcal{A}_l^{r_l}],\label{a2}\\
\tr[\Delta^{\otimes2}\mathbb{A}^{r_l}_l] &= 0. \label{a3}
\end{align}   
\end{subequations}
Since $\Delta$ is Hermitian and traceless, it can be expressed as $\Delta=\sum_ia_i\sigma_i$ with $a_i\in \mathbb{R},~\text{for}~ i \in \{x,y,z\}$, and $\sigma$'s being the Pauli operators. Now, substituting the $\Delta=\sum_ia_i\sigma_i$ in Eqs.(\ref{a3}) \& (\ref{a2}) and comparing same degree coefficients on both sides we respectively obtain
\begin{subequations}
\begin{align}
\tr[(\sigma_i\otimes\sigma_j)\mathbb{A}^{r_l}_l]&=0, ~\forall~i,j\in\{x,y,z\},\label{c1}\\
\tr[(\sigma_i\otimes\id-\id\otimes\sigma_i)\mathbb{A}^{r_l}_l]&=2 \tr[\sigma_i\mathcal{A}_l^{r_l}],~\forall~i\in\{x,y,z\}.\label{c2}
\end{align}    
\end{subequations}
Given a particular (but arbitrary) choice of $l$ and $r_l$, the operator $\mathbb{A}^{r_l}_l$ can be expressed in Hilbert-Schmidt form as 
\begin{align}
\mathbb{A}^{r_l}_l=b\id^{\otimes2}+\sum_ic_i\sigma_i\otimes\id+\sum_id_i\id\otimes\sigma_i+\sum_{i,j}e_{ij}\sigma_i\otimes\sigma_j.\label{hs} 
\end{align}
Since, $\Pi_{\bf r}=(\mathrm{F}\otimes\mathrm{F})\circ\text{Swap}(\Pi_{\bf r})~\forall~{\bf r}$, we thus have $\mathbb{A}^{r_l}_l=(\mathrm{F}\otimes\mathrm{F})\circ\text{Swap}\left(\mathbb{A}^{r_l}_l\right)$, implying $c_i=-d_i$. Now, from Eqs.(\ref{a1}),~(\ref{c1}),~(\ref{c2}),~\&~(\ref{hs}) we obtain
\begin{align}
b=\frac{1}{2}\tr[\mathcal{A}_l^{r_l}],~~c_i=\frac{1}{4}\tr[\sigma_i\mathcal{A}_l^{r_l}],~~e_{ij}=0,~\forall~i,j\in\{x,y,z\}.
\end{align}
Thus we have
\begin{align}
\mathbb{A}^{r_l}_l&=\frac{1}{2}\tr[\mathcal{A}_l^{r_l}]\id^{\otimes2}+\sum_i\frac{1}{4}\tr[\sigma_i\mathcal{A}_l^{r_l}]\sigma_i\otimes\id-\sum_i\frac{1}{4}\tr[\sigma_i\mathcal{A}_l^{r_l}]\id\otimes\sigma_i,\nonumber\\
&=\frac{1}{2}\big(\frac{1}{2}\tr[\mathcal{A}_l^{r_l}]\id+\sum_i\frac{1}{2}\tr[\sigma_i\mathcal{A}_l^{r_l}]\sigma_i\big)\otimes\id+\frac{1}{2}\id\otimes\big(\frac{1}{2}\tr[\mathcal{A}_l^{r_l}]\id-\sum_i\frac{1}{2}\tr[\sigma_i\mathcal{A}_l^{r_l}]\sigma_i\big),\nonumber\\ 
&=\frac{1}{2}\left(\mathcal{A}_l^{r_l}\otimes\id+\id\otimes\mathrm{F}\big(\mathcal{A}_l^{r_l}\big)\right).
\end{align}
This completes the proof. 
\end{proof}

\section{Enhanced measurement compatibility in GPT}

\noindent {\bf Framework of GPT.--} Originally developed in the 1960's \cite{Ludwig1967,Ludwig1968,Mielnik1968,Mackey1977} and more recently revitalized \cite{Hardy2001,Barrett2007,Chiribella2011}—offers valuable mathematical insight. In particular, assuming a quantum description for individual systems, a wide range of mathematically consistent composite system descriptions—bounded by the minimal and maximal tensor products—are possible under the constraints of no-signaling and tomographic locality \cite{Namioka1969,Klay1987,Barnum2010,Torre2012,Naik2022,Patra2023}. Among these, the minimal composition offers deeper insight into the extended range of admissible sharpness values for joint measurability observed in the configurations of \textbf{Corollary 1} of main manuscript. 

\noindent$\bullet$~{\it Elementary System:} An elementary system of a GPT is specified by the triple $(\Omega,\mathbb{E},\mathbb{T})$. The set $\Omega$ denotes the normalized state space, which is generally assumed to be a convex compact set embedded in some real vector space. Mathematically, often it is easy to work with the set of Unnormalized states which form a convex cone. The set $\mathbb{E}$ corresponds to the effect space, where an effect $e\in\mathbb{E}$ yields probability $e(\omega)$ on the state $\omega\in\Omega$. A measurement $M$ is a collection of effects adding to unit effect, i.e. $M\equiv\{e_k~|~\sum_ke_k=u\}$, where the unite effect is defined as $u(\omega)=1~\forall~\omega\in\Omega$. Finally, the set $\mathbb{T}$ denotes the reversible transformations allowed on the system. For instance, a qubit in the language of GPT is described as follows:
\begin{align}
\begin{rcases}
\Omega\equiv\mathcal{D}(\mathbb{C}^2):=\left\{\rho\in\mathcal{L}(\mathbb{C}^2)~|~\rho\ge0~\&\tr[\rho]=1\right\},\\
\mathbb{E}\equiv\mathcal{E}(\mathbb{C}^2):=\left\{\pi\in\mathcal{L}(\mathbb{C}^2)~|~0\le\pi\le\mathbf{I}_2\right\},\\
\mathbb{T}\equiv\mathcal{U}(\mathbb{C}^2):=\left\{U\in\mathcal{L}(\mathbb{C}^2)~|~UU^\dagger=U^\dagger U=\mathbf{I}_2\right\}
\end{rcases},
\end{align}
where $\mathcal{U}(\cdot)$ denotes the set of unitary operators. 

\noindent$\bullet$~{\it Composite System:} A GPT also specify the composition rules constrained by no-signaling principle, that prohibits any superluminal communication. Under the assumption of tomographic locality—where global states are fully specified by local measurements \cite{Hardy2011}—composite systems lie between the minimal and maximal tensor products \cite{Namioka1969}. For instance, for two elementary qubits, respectively denoted by A and B, the minimal and maximal compositions are defines as 
\begin{subequations}
\begin{align}
\begin{rcases}
\Omega^{AB}_{\min}&:=\left\{\mathrm{Ch}\{\rho_{_A}\otimes\rho_{_B}\}~|~\rho_{_A}\in\mathcal{D}(\mathbb{C}^2_A)~\&~\rho_{_B}\in\mathcal{D}(\mathbb{C}^2_B)\right\},\\
\mathbb{E}^{AB}_{\min}&:=\left\{\pi_{_{AB}}\in\mathcal{L}(\mathbb{C}^2_A\otimes\mathbb{C}^2_B),~\text{s.t.} ~0\le\tr[\pi_{_{AB}}\rho_{_{AB}}]\le1,~\forall~\rho_{_{AB}}\in\Omega^{AB}_{\min}\right\},\\
\mathbb{T}^{AB}_{\min}&:=\left\{U_A\otimes U_B~|~U_A\in\mathcal{U}(\mathbb{C}^2_A),~U_B\in\mathcal{U}(\mathbb{C}^2_B)\right\}
\end{rcases},\\
\begin{rcases}
\mathbb{E}^{AB}_{\max}&:=\left\{\mathrm{Ch}\{\pi_{_A}\otimes\pi_{_B}\}~|~\pi_{_A}\in\mathcal{E}(\mathbb{C}^2_A)~\&~\pi_{_B}\in\mathcal{E}(\mathbb{C}^2_B)\right\},\\
\Omega^{AB}_{\max}&:=\left\{\rho_{_{AB}}\in\mathcal{L}(\mathbb{C}^2_A\otimes\mathbb{C}^2_B),~\text{s.t.}~0\le\tr[\pi_{_{AB}}\rho_{_{AB}}\le1,~\forall~\pi_{_{AB}}\in\mathbb{E}^{AB}\right\},~~~\\
\mathbb{T}^{AB}_{\max}&:=\left\{U_A\otimes U_B~|~U_A\in\mathcal{U}(\mathbb{C}^2_A),~U_B\in\mathcal{U}(\mathbb{C}^2_B)\right\}
\end{rcases},
\end{align}
\end{subequations}
where $\mathrm{Ch}\{\cdot\}$ denotes convex hull of a set. While minimal composition allows only separable states, the maximal one allows effects that are separable only. The standard quantum composition strictly lies in between,
\begin{align}
\begin{rcases}
\Omega^{AB}_{Q}:=\mathcal{D}(\mathbb{C}^2_A\otimes\mathbb{C}^2_B)\equiv\left\{\rho_{_{AB}}\in\mathcal{L}(\mathbb{C}^2_A\otimes\mathbb{C}^2_B),~\text{s.t.}~\rho_{_{AB}}\ge0~\&\tr[\rho_{_{AB}}]=1\right\},\\
\mathbb{E}^{AB}_Q:=\mathcal{E}(\mathbb{C}^2_A\otimes\mathbb{C}^2_B)\equiv\left\{\pi_{_{AB}}\in\mathcal{L}(\mathbb{C}^2_A\otimes\mathbb{C}^2_B)~|~0\le\pi_{_{AB}}\le\mathbf{I}_4\right\},\\
\mathbb{T}^{AB}_Q:=\mathcal{U}(\mathbb{C}^2_A\otimes\mathbb{C}^2_B)\equiv\left\{U_{_{AB}}\in\mathcal{L}(\mathbb{C}^2_A\otimes\mathbb{C}^2_B)~|~U_{_{AB}}U^\dagger_{_{AB}}=U^\dagger_{_{AB}} U_{_{AB}}=\mathbf{I}_4\right\}    
\end{rcases}.
\end{align}
Notably, maximal Composition includes bipartite states that are not permitted in quantum theory, while minimal Composition allows bipartite effects that are not allowed in quantum composition. 

\noindent$\bullet$~{\it Measurement compatibility in minimal composition:} As noted above, although the state space of minimal composition is restricted, the effect space is enlarged compared to standard quantum theory. This leads to enhanced compatibility for certain measurements, as shown below.
\begin{manualtheorem}{1S}\label{theo1s}
The observables \(X^\lambda, Y^\lambda,\) and \(Z^\lambda\) are 2-copy compatible for all \(\lambda \in [0,1]\) in the minimal tensor product GPT.
\end{manualtheorem}
\begin{proof}
Consider the set of operators \(\{\Pi^{\#}_{^{[a,b,c]}}~|~a,b,c = \pm1\} \subset \mathcal{L}(\mathbb{C}^2 \otimes \mathbb{C}^2)\), defined as  
\begin{align}\label{pipopt}
\Pi^{\#}_{^{[a,b,c]}}&:=\frac{1}{16}\Big(2~\id^{\otimes 2} + a\dacb{X,\id}~+ b\dacb{Y,\id}~+ c\dacb{Z,\id}~\nonumber\\
&\hspace{2cm}+ ab\dacb{X,Y}~ + bc\dacb{Y,Z}~ + ca\dacb{Z,X}~\Big).
\end{align}
Although these operators are not positive and thus cannot be valid effects in quantum theory, they add to the identity operator, i.e. 
\begin{align}
\sum_{a,b,c}\Pi^{\#}_{^{[a,b,c]}} = \id\otimes\id.    
\end{align}
Furthermore, on any qubit state $\rho_{\Vec{m}}\in\mathcal{D}(\mathbb{C}^2)$ and for $a,b,c=\pm1$, they satisfy the following identities
\begin{subequations}
\begin{align}
\sum_{b,c=\pm1}\tr\left(\Pi^{\#}_{^{[a,b,c]}}\rho^{\otimes2}_{\Vec{m}}\right)&=\tr\left(\mathrm{P}^a_{\hat{x}}\rho_{\Vec{m}}\right),\\
\sum_{a,c=\pm1}\tr\left(\Pi^{\#}_{^{[a,b,c]}}\rho^{\otimes2}_{\Vec{m}}\right)&=\tr\left(\mathrm{P}^b_{\hat{y}}\rho_{\Vec{m}}\right),\\
\sum_{a,b=\pm1}\tr\left(\Pi^{\#}_{^{[a,b,c]}}\rho^{\otimes2}_{\Vec{m}}\right)&=\tr\left(\mathrm{P}^c_{\hat{z}}\rho_{\Vec{m}}\right),
\end{align}    
\end{subequations}
thereby reproducing the measurement statistics of all three Pauli observables. We will now show that these operators are valid effects in minimal composition, i.e, $\Pi^{\#}_{^{[a,b,c]}}$ will yield positive probabilities on all two qubit separable states for all $a,b,c$. To establish this it is sufficient to show that it yields positive probabilities on all product states of the form $\rho_{\vec{r}}\otimes\rho_{\vec{s}}$, where $\vec{r}$ and $\vec{s}$ denote the respective Block vectors. 

The spin-flip map $\mathrm{F}:\mathcal{L}(\mathbb{C}^2)\mapsto\mathcal{L}(\mathbb{C}^2)$ can be written as  $\mathrm{F}= T\circ\sigma_y$, where $\sigma_y$ is the unitary map corresponding to Pauli $Y$ and $T$ is the transpose map in computational basis. From the fact that both $\sigma_y$ and $T$ is inner product preserving it follows that
\begin{align}\label{flip1}
\tr(A^\dagger B)= \tr (\mathrm{F}(A^\dagger)\mathrm{F}(B))), ~\forall~A,B\in\mathcal{L}(\mathcal{H}).
\end{align}

One can readily check that 
\begin{align}
\Pi^\#_{[a,b,c]}=(\id\otimes \mathrm{F})(\Pi^{\uparrow\downarrow}_{[a,b,c]})~\forall~a,b,c.    
\end{align}
Since, $\Pi^{\uparrow\downarrow}_{[a,b,c]}$'s are positive we have, for all $a,b,c$, we have 
\begin{align*}
&\tr \left(\Pi^{\uparrow\downarrow}_{[a,b,c]}\rho_{\vec{r}}\otimes\rho_{\vec{s}}\right)\ge 0,\quad \forall~\rho_{\vec r},\rho_{\vec s}\in\mathcal{D}(\mathbb{C}^2)\\
&\tr\left(\id \otimes \mathrm{F}( \Pi^{\uparrow\downarrow}_{[a,b,c]})\id \otimes \mathrm{F}(\rho_{\vec{r}}\otimes\rho_{\vec{s}})\right)\ge 0,\quad \forall~\rho_{\vec r},\rho_{\vec s}\in\mathcal{D}(\mathbb{C}^2)\\
&\tr\left(\Pi^{\#}_{[a,b,c]}\rho_{\vec{r}}\otimes\rho_{-\vec{s}}\right)\ge 0\quad \forall~\rho_{\vec r},\rho_{\vec s}\in\mathcal{D}(\mathbb{C}^2),
\end{align*}
This completes the proof.
\end{proof}

\noindent Naturally the question arises whether a result analogous to Theorem \ref{theo1s} holds for the set of observables $\mathrm{SIC}_4 \equiv \{\sigma_{\hat{n}_r}\}_{r=0}^3$ analyzed in \textbf{Theorem 2} of the main manuscript. We answer this question affirmatively by proving an even more generic result for arbitrary POVM measurements instead of spin observables. Consider $N$ qubit POVMs $\{\mathcal{A}_{l}\equiv\{\mathcal{A}_{l}^{r_l}\}_{r_l=1}^{K_l}\}_{l=1}^N$ each consisting of multiple (possibly different) effects. Recalling Definition \ref{def1s} for antiparallel configuration, the set will be jointly measurable if there exist a POVM 
\begin{align}
\mathcal{G}\equiv\left\{\Pi_{\bf r}\in\mathcal{E}((\mathbb{C}^2)^{\otimes2})~|~\sum_{\bf r}\Pi_{\bf r}=\id^{\otimes2}\right\}~\text{with}~{\bf r}:=[r_1,\cdots,r_N]\in K_1\times\cdots\times K_N,\label{jm}
\end{align}
such that for all $\rho_{\vec{m}}\in\mathcal{D}(\mathbb{C}^2)$
\begin{align}
\sum_{{\bf r}\setminus r_l}\Tr\left[\Pi_{\bf r}\left(\rho_{\vec{m}}\otimes \mathrm{F}(\rho_{\vec{m}})\right)\right]=\Tr\left[\mathcal{A}_{l}^{r_l}\rho_{\vec{m}}\right],~~ \forall~l\in\{1,\cdots,N\},~\&~r_l\in K_l.\label{reproduceOk}
\end{align}    
With this we are now in a position to prove the \textbf{Theorem 4} of main manuscript.
\begin{proof}
Let $N$ qubit observables $\{\mathcal{A}_l\}_{l=1}^N$ be compatible in quantum theory on the {\footnotesize$\uparrow\hspace{-.05cm}\downarrow$} configuration. Thus there exists a POVM (\ref{jm}) satisfying Eq.(\ref{reproduceOk}). Using the fact that map $\mathrm{F}$ is its own dual we can rewrite Eq.(\ref{reproduceOk}) as
\begin{align}
\sum_{{\bf r}\setminus r_l}\Tr\left[\text{id}_2\otimes\mathrm{F}\left(\Pi_{\bf r}\right)\left(\rho_{\vec{m}}\otimes \rho_{\vec{m}}\right)\right]=\Tr\left[\mathcal{A}_{l}^{r_l}\rho_{\vec{m}}\right],~~ \forall~l\in\{1,\cdots,N\},~r_l\in K_l,~\&~\rho_{\vec{m}}.\label{reproduceOk1}
\end{align}   
In general the operators $\Pi_{\bf r}^{\#}:=\text{id}_2\otimes\mathrm{F}\left(\Pi_{\bf r}\right)$ are not positive, but nevertheless they yield positive probabilities on separable states $\sigma_{AB}\in\Omega^{AB}_{\min}$. This follows from the fact that
\begin{align}
\Tr\left[\Pi^{\#}_{\bf r}\sigma_{AB}\right]=\Tr\left[\Pi_{\bf r}\left(\text{id}_2\otimes \mathrm{F}(\sigma_{AB})\right)\right]\geq0,~\forall~\sigma_{AB}\in\Omega^{AB}_{\min}.    
\end{align}
Furthermore, $\mathrm{F}$ being unital implies
\begin{align}
\sum_{\bf r}\Pi^{\#}_{\bf r}=\sum_{\bf r}\text{id}_2\otimes \mathrm{F}\left(\Pi_{\bf r}\right)=\text{id}_2\otimes \mathrm{F}\left(\sum_{\bf r}\Pi_{\bf r}\right)=\text{id}_2\otimes \mathrm{F}\left(\id^{\otimes2}\right)=\id^{\otimes2}.
\end{align}
Thus the collection of operators $\mathcal{A}^{\#}\equiv\left\{\Pi^{\#}_{\bf r}\right\}_{\bf r}$ constitutes a valid measurement in the minimal tensor product GPT. On the other hand, Eq.(\ref{reproduceOk1}) ensures that this collection of operators also reproduce the statistics of $\{\mathcal{A}_l\}_{l=1}^N$ on the {\footnotesize$\uparrow\hspace{-.05cm}\uparrow$} configuration. This completes the proof.
\end{proof}
\begin{manualremark}{1S}
In \textbf{Theorem $2$} of our main manuscript we have shown that the observables $\mathrm{SIC}_4$ are compatible in quantum theory on {\footnotesize$\uparrow\hspace{-.05cm}\downarrow$} configuration. Now as a corollary of \textbf{Theorem 4} it follows that these observables are compatible on {\footnotesize$\uparrow\hspace{-.05cm}\uparrow$} configuration in the minimal tensor product GPT.
\end{manualremark}

\begin{manualremark}{2S}
One might expect that the converse of \textbf{Theorem 4} also holds true, namely joint measurability of set of qubit POVMs on {\footnotesize$\uparrow\hspace{-.05cm}\uparrow$} configuration in the minimal tensor product GPT would imply  their joint measurability on {\footnotesize$\uparrow\hspace{-.05cm}\downarrow$} configuration in quantum theory. However, as we argue this is not the case in general. Joint measurability of set of qubit POVMs on {\footnotesize$\uparrow\hspace{-.05cm}\uparrow$} configuration in the minimal tensor product GPT implies existence of a collection of operators $\left\{\Pi_{\bf r}\in\mathbb{E}^{AB}_{\min}~|~\sum_{\bf r}\Pi_{\bf r}=\id^{\otimes2}\right\}_{\bf r}$, where some of the $\Pi_{\bf r}$'s could be entangled. Consequently, the respective $\text{id}_2\otimes\mathrm{F}\left(\Pi_{\bf r}\right)$'s are not positive operators, and hence do not constitute a valid measurement in quantum theory. 
\end{manualremark}

\section{Mean-King problem and Bub's Cryptography}
\noindent In the main manuscript we have analyzed the VAA's strategy for mean King retrodiction task when Bob chooses his measurement from the set $\{X,Y,Z\}$.  In this section, we extend the scenario to four observables, where Bob chooses his measurement from the set $\mathrm{SIC}_4 \equiv \{\sigma_{\hat{n}_r}\}_{r=0}^3$, where {\footnotesize$\hat{n}_r := \frac{(-1)^{\bar{\delta}_{r,0}}}{\sqrt{3}}\big((-1)^{\delta_{r,1}}, (-1)^{\delta_{r,2}}, (-1)^{\delta_{r,3}}\big)$}. Here also Alice shares the channel part of the singlet state with Bob, who sends the updated  state back to Alice after performing the measurement of his choice. Alice performs the POVM $\mathcal{G}^{\uparrow\hspace{-.05cm}\downarrow}_{\mathrm{SIC}_4}=\{ \Pi^{_{\uparrow\hspace{-.05cm}\downarrow}}_{[\Gamma,a]}~|~a = \pm1,~\&~\Gamma\in\{X,Y,Z\}\}$, with 
\begin{align}
\Pi^{_{\uparrow\hspace{-.05cm}\downarrow}}_{[\Gamma,a]}=\frac{1}{12} \Bigl(2 \id^{\otimes 2} -2\Gamma^{\otimes2}
+a\sqrt{3}\dcb{\Gamma,\id}~ + \sum_{\Omega\in\{\Gamma\}^{\footnotesize\complement}}\Omega^{\otimes2}\Bigr),
\end{align}
on the ancilla and channel particle. Depending upon the measurement outcome, she perfectly retrodicts Bob's measurement outcome as per Table \ref{tab2}.
\begin{table}[t!]
\centering
\begin{tabular}{|c||c|c|c|c|c|}
\hline
  &~~ Outcome $\backslash$ Retrodiction ~~& \cellcolor{blue!35}~~~~~~$\sigma_{\hat{n}_0}$ ~~~~~~ & \cellcolor{blue!35}~~~~~~$\sigma_{\hat{n}_1}$~~~~~~ & ~~~~~~$\sigma_{\hat{n}_2}$~~~~~~ & ~~~~~~$\sigma_{\hat{n}_3}$~~~~~~\\ \hline\hline
\multirow{6}{*}{$\mathcal{G}^{\uparrow\hspace{-.05cm}\downarrow}_{\mathrm{SIC}_4}$} 
  & $\Pi^{_{\uparrow\hspace{-.05cm}\downarrow}}_{[X,+1]}$ & \cellcolor{green!50}$-1$ & \cellcolor{green!50}$-1$ &$+1$ &$+1$\\ \cline{2-6}
  & $\Pi^{_{\uparrow\hspace{-.05cm}\downarrow}}_{[X,-1]}$ & \cellcolor{green!50}$+1$ & \cellcolor{green!50}$+1$ & $-1$ &$-1$ \\ \cline{2-6}
  & $\Pi^{_{\uparrow\hspace{-.05cm}\downarrow}}_{[Y,+1]}$ & \cellcolor{red!50}$-1$ & \cellcolor{red!50}$+1$ & $-1$ &$+1$\\ \cline{2-6}
  & $\Pi^{_{\uparrow\hspace{-.05cm}\downarrow}}_{[Y,-1]}$ & \cellcolor{red!50}$+1$ & \cellcolor{red!50}$-1$ & $+1$ &$-1$\\ \cline{2-6}
  & $\Pi^{_{\uparrow\hspace{-.05cm}\downarrow}}_{[Z,+1]}$ & \cellcolor{red!50}$-1$ &\cellcolor{red!50} $+1$ & $+1$ &$-1$\\ \cline{2-6}
  & $\Pi^{_{\uparrow\hspace{-.05cm}\downarrow}}_{[Z,-1]}$ & \cellcolor{red!50}$+1$ & \cellcolor{red!50}$-1$ & $-1$ &$+1$\\ \hline
\end{tabular}
\caption{(Color online)  Bob performs any one of the measurements from the set $\mathrm{SIC}_4 \equiv \{\sigma_{\hat{n}_r}\}_{r=0}^3$, where {\footnotesize$\hat{n}_r := \frac{(-1)^{\bar{\delta}_{r,0}}}{\sqrt{3}}\big((-1)^{\delta_{r,1}}, (-1)^{\delta_{r,2}}, (-1)^{\delta_{r,3}}\big)$}. To win the mean King’s challenge, Alice performs the measurement $\mathcal{G}^{\uparrow\hspace{-.05cm}\downarrow}_{\mathrm{SIC}_4}=\{ \Pi^{_{\uparrow\hspace{-.05cm}\downarrow}}_{[\Gamma,a]}~|~a = \pm1,~\&~\Gamma\in\{X,Y,Z\}\}$ on her ancilla and the channel particle (received from Bob). Upon obtaining the outcome, she retrodicts Bob’s outcomes for four observables $\mathrm{SIC}_4$ as per the table. When Bob restricts his measurements to $\sigma_{\hat{n}_0}$ and $\sigma_{\hat{n}_1}$, a secure key can be established: outcomes corresponding to $\Pi^{_{\uparrow\hspace{-.05cm}\downarrow}}_{[Y/z,\pm1]}$  (marked in red) are used for detection of possible eavesdropping, while the correlation in the remaining outcomes (marked in green) generates a raw key.}\label{tab2}
\vspace{-.35cm}
\end{table}

\noindent{\bf Bub's QKD protocol.--} The Quantum Key Distribution (QKD) protocol proposed by Bub in also possible in this case if Bob's measurement choice is limited between any two observables from the set $\mathrm{SIC}_4$. For instance, let Bob randomly chooses to perform one of the measurements $\sigma_{\hat{n}_0}$, $\sigma_{\hat{n}_1}$. Now, if Alice perform the POVM $\mathcal{G}^{\uparrow\hspace{-.05cm}\downarrow}_{\mathrm{SIC}_4}$, then they can detect the possible interference of Eve by Alice publicly announcing the outcomes whenever one of the projector $\Pi^{_{\uparrow\hspace{-.05cm}\downarrow}}_{[Y/Z,\pm1]}$ click. In case of no discrepancy they proceed with the QKD protocol, and use the outcomes $\Pi^{_{\uparrow\hspace{-.05cm}\downarrow}}_{[X,\pm1]}$ to generate shared random key (see Table \ref{tab2}).

\subsection{Connection between the POVM $\mathcal{G}^{_{\uparrow\hspace{-.05cm}\downarrow}}_{\mathrm{MUB}}$ and EJM} \noindent All the effects \(\Pi^{_{\uparrow\hspace{-.05cm}\downarrow}}_{^{[a,b,c]}}\)'s are rank one, and can be expressed as \(\Pi^{_{\uparrow\hspace{-.05cm}\downarrow}}_{^{[a,b,c]}}\propto\xi_{[a,b,c]}\), where \(\xi:=\ket{\xi}\bra{\xi}\) with 
\begin{align}
\ket{\xi_{[a,b,c]}}:= \frac{1}{2}(\ket{\psi^-}-a\ket{\phi^-}+\mathbf{i}~b\ket{\phi^+}+c\ket{\psi^+}),
\end{align}
where \(\ket{\phi^\pm}=\frac{1}{2}(\ket{00}\pm\ket{11}),~\ket{\psi^\pm}=\frac{1}{2}(\ket{01}\pm\ket{10}),~\&~\mathbf{i}=\sqrt{-1}\).
\begin{figure}[t!]
\centering
\subfloat[$\{\xi_1,\xi_2,\xi_4,\xi_7\}$]{\includegraphics[width=0.30\linewidth]{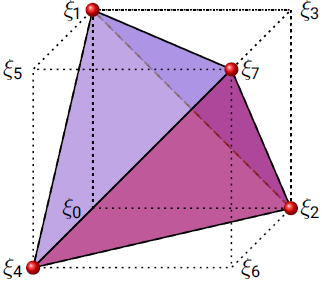}}
\hspace{0.1cm}
\subfloat[$\{\xi_l\}_{l=0}^7$]{\includegraphics[width=0.30\linewidth]{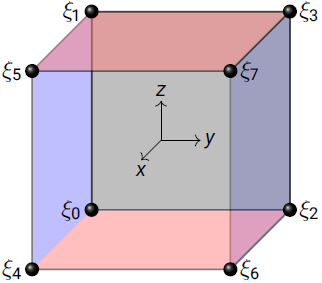}}
\hspace{0.1cm}
\subfloat[$\{\xi_0,\xi_3,\xi_5,\xi_6\}$]{\includegraphics[width=0.30\linewidth]{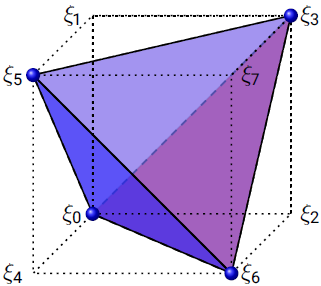}}
\caption{(Color online) The vectors $\{\xi_l\}_{l=0}^7$ (middle). Each vertex is orthogonal to the vertices connected through face diagonal, namely the vectors \(\{\xi_1,\xi_2,\xi_4,\xi_7\}\) are mutually orthogonal (left) and the vectors \(\{\xi_0,\xi_3,\xi_5,\xi_6\}\) are mutually orthogonal (right).\vspace{-.5cm}}\label{Figs}
\end{figure}

Representing \(\xi_{[a,b,c]}\equiv\xi_l\), with \(l:=2(a+1)+(b+1)+(c+1)/2\), when these vectors are depicted as the vertices of a cube, each vertex becomes orthogonal the vertices connected through face diagonal (see Fig. \ref{Figs}). 
If we restrict the outcomes with positive product i.e. $abc=+1$, then these four vectors $\{\ket{\xi_{[a,b,c]}}\}$ are precisely the eigenvectors of the elegant joint measurement (EJM) \cite{Gisin2019}:
\begin{align}
\ket{\xi_{[a,b,c]}}= \sqrt{\frac{3}{2}}\ket{\vec{m}_{abc},-\vec{m}_{abc}}+\mathbf{i}\frac{\sqrt{3}-1}{2}\ket{\psi^-},\label{EJM}
\end{align}
where $\ket{\vec{m}_{abc}}$'s are the qubit states corresponding to the Bloch vectors $\vec{m}_{abc}=\frac{1}{\sqrt{3}}(a,b,c)$ forming the vertices of a regular tetrahedron. The remaining four outcomes with \(abc = -1\) form another EJM, which we call the inverted-EJM as the Bloch vectors \(\vec{m}_{abc}\) form the inverted tetrahedron as that of the \(abc = +1\) case. Therefore, performing regular-EJM or inverted-EJM based on the outcome of an unbiased coin toss realizes the measurement $\mathcal{G}^{_{\uparrow\hspace{-.05cm}\downarrow}}_{\mathrm{MUB}}$.

\section{Joint measurability of Three Symmetric Spin Observables on Antiparallel configuration}
\begin{figure}[h!]
\centering
\includegraphics[height=8cm,width=11cm]{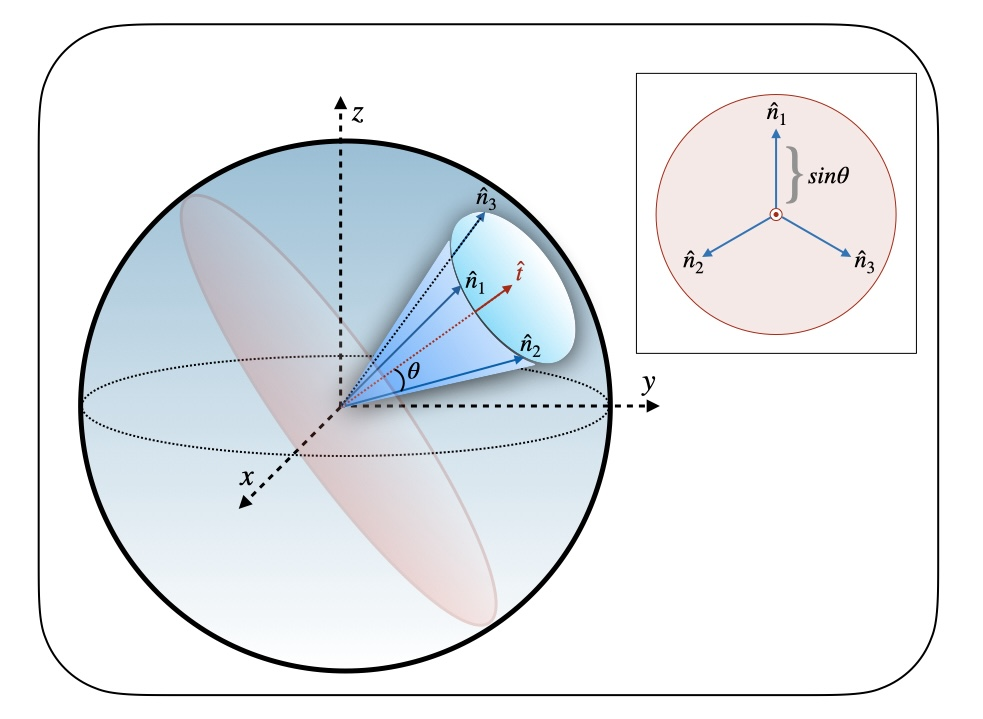}
\caption{(Color online) The observables $\left\{\sigma_{\hat{n}_i}\right\}_{i=1}^3$ are symmetrically chosen by fixing the relative angles between any two of them to be identical. Each of them makes an angle $\theta$ with $\hat{t}=\tfrac{1}{\sqrt{3}}(1,1,1)^{\mathrm{T}}$. Inset: view from the top of the $\hat{t}$ vector. \vspace{-.5cm}} \label{Figs0}
\end{figure}

\noindent In this section we analyze the joint measurability  of a set of three spin observables along directions $\hat{n}_1, \hat{n}_2,$ and $\hat{n}_3$ that are symmetrically chosen by fixing the relative angles between any two of the spin directions are identical, i.e., 
\begin{align}
 \hat{n}_1\cdot\hat{n}_2 = \hat{n}_2\cdot\hat{n}_3 = \hat{n}_3\cdot\hat{n}_1~.   
\end{align}
Without loss of any generality, by choosing $\hat{t}=\tfrac{1}{\sqrt{3}}(1,1,1)^{\mathrm{T}}$, the above symmetry can be ensured by fixing 
\begin{subequations}
\begin{align}
&\hat{n}_1=(\alpha,\beta,\beta)^{\mathrm{T}}, \quad\hat{n}_2=(\beta,\alpha,\beta)^{\mathrm{T}}, \quad\hat{n}_3=(\beta,\beta,\alpha)^{\mathrm{T}},~\text{where}\\
&\alpha:=\tfrac{1}{3}\big(\sqrt{3}\cos\theta+\sqrt{6}\sin\theta\big),\quad\beta:=\tfrac{1}{6}\big(2\sqrt{3}\cos\theta-\sqrt{6}\sin\theta\big),
\end{align} \label{sym-triple} 
\end{subequations}
with $\theta:=\cos^{-1}(\hat{t}\cdot\hat{n}_i)\in[0,\pi/2]$ (see Fig.~\ref{Figs0}). The case $\theta=0$ is not interesting as all the observables become identical ($\hat{n}_1=\hat{n}_2=\hat{n}_3=\hat{t}$), and hence are trivially jointly measurable. This symmetric configuration imposes a corresponding symmetry on the parent POVM structure as stated in the following Lemma.
\begin{manuallemma}{1S}\label{lem1}
If the symmetric spin observables $\{\sigma_{\hat{n}_i}\}_{i=1}^3$, with $\hat{n}_i$'s defined as in Eqs.(\ref{sym-triple}), are jointly measurable on antiparallel configuration, then there exists a parent POVM $\mathcal{G}^{_{\uparrow\hspace{-.05cm}\downarrow,\theta}}\equiv\left\{\Pi^{_{\uparrow\hspace{-.05cm}\downarrow,\theta}}_{{[a,b,c]}}:=\Pi^{\theta}_{{[a,b,c]}}\in\mathcal{E}\left(\left(\mathbb{C}^2\right)^{\otimes2}\right) ~|~\sum_{[a,b,c]} \Pi^{\theta}_{[a,b,c]} = \id^{\otimes 2}\right\}$ satisfying the following conditions (for all $a,b,c\in\{\pm1\}$):
\begin{itemize}
\item[{\bf (1)}] $\operatorname{Rank}\left[\Pi^{\theta}_{[a,b,c]}\right]=1$~;
\item[{\bf (2)}] $\Pi^{\theta}_{[-a,-b,-c]}=\mathrm{F}\otimes\mathrm{F}\left(\Pi^{\theta}_{[a,b,c]}]\right)$~;
\item[{\bf (3)}] $\Pi^{\theta}_{[a,b,c]}=\mathrm{R_{XY}}\otimes\mathrm{R_{XY}}\left(\Pi^{\theta}_{[b,a,c]}\right)=\mathrm{R_{YZ}}\otimes\mathrm{R_{YZ}}\left(\Pi^{\theta}_{[a,c,b]}\right)=\mathrm{R_{ZX}}\otimes\mathrm{R_{ZX}}\left(\Pi^{\theta}_{[c,b,a]}\right)$, where $\mathrm{R_{XY}}$ is the anti unitary operation satisfying $\mathrm{R_{XY}}(\id)=\id,~\mathrm{R_{XY}}(X)=Y,~\mathrm{R_{XY}}(Y)=X$ and $\mathrm{R_{XY}}(Z)=Z$. $\mathrm{R_{YZ}}$ and $\mathrm{R_{ZX}}$ are defined analogously.
\end{itemize}
\end{manuallemma}
\begin{proof}
~\\
\noindent{\bf (1)} The requirement that $\mathcal{G}^{_{\uparrow\hspace{-.05cm}\downarrow,\theta}}$ should reproduces outcome statistics of the observables $\{\sigma_{\hat{n}_i}\}_{i=1}^3$ demands
\begin{align}
\begin{rcases}
&\sum_{b,c}\tr\left[\Pi^{\theta}_{[a,b,c]}\mathrm{P}^{-a}_{\hat{n}_1}\otimes\mathrm{P}^{a}_{\hat{n}_1}\right]=0,~\forall~a; \quad\quad\sum_{a,c}\tr\left[\Pi^{\theta}_{[a,b,c]}\mathrm{P}^{-b}_{\hat{n}_2}\otimes\mathrm{P}^{b}_{\hat{n}_2}\right]=0,~\forall~b;\\
&\hspace{4cm}\sum_{a,b}\tr\left[\Pi^{\theta}_{[a,b,c]}\mathrm{P}^{-c}_{\hat{n}_3}\otimes\mathrm{P}^{c}_{\hat{n}_3}\right]=0,~\forall~c;
\end{rcases}\label{rank1}
\end{align}
where $\mathrm{P}^k_{\hat{n}}:=\ket{\psi^k_{\hat{n}}}\bra{\psi^k_{\hat{n}}}$ be the eigen-projector of the observable $\sigma_{\hat{n}}$ with eigenvalue $k\in\{\pm1\}$. Eq.(\ref{rank1}) further implies that $\forall~a,b,c$
\begin{align}\label{ortho}
\tr\left[\Pi^{\theta}_{[a,b,c]}\mathrm{P}^{-a}_{\hat{n}_1}\otimes\mathrm{P}^{a}_{\hat{n}_1}\right]=0,~~
\tr\left[\Pi^{\theta}_{[a,b,c]}\mathrm{P}^{-b}_{\hat{n}_2}\otimes\mathrm{P}^{b}_{\hat{n}_2}\right]=0,~~
\tr\left[\Pi^{\theta}_{[a,b,c]}\mathrm{P}^{-c}_{\hat{n}_3}\otimes\mathrm{P}^{c}_{\hat{n}_3}\right]=0.    
\end{align}
For every fixed choices of the triple $[a,b,c]$, the vectors $\left\{\ket{\psi^{-a}_{\hat{n}_1}}\otimes\ket{\psi^{a}_{\hat{n}_1}},~\ket{\psi^{-b}_{\hat{n}_2}}\otimes\ket{\psi^{b}_{\hat{n}_2}},~\ket{\psi^{-c}_{\hat{n}_3}}\otimes\ket{\psi^{c}_{\hat{n}_3}}\right\}$ span a three dimensional subspace of $\mathbb{C}^2\otimes\mathbb{C}^2$. Therefore, the operator $\Pi^{\theta}_{[a,b,c]}$ is proportional to the orthogonal one dimensional subspace and hence is of Rank-$1$.

\noindent{\bf (2)} Let the POVM $\left\{\Pi^{\prime~\theta}_{{[a,b,c]}}\right\}$ reproduces measurement statistics of $\left\{\sigma_{\hat{n}_i}\right\}_{i=1}^3$ for a state $\rho$ on the configuration $\rho\otimes\mathrm{F}(\rho)$. Manifestly, the POVM $\left\{\Pi^{\prime\prime~\theta}_{{[a,b,c]}}:=\Pi^{\prime~\theta}_{{[-a,-b,-c]}}\right\}$ reproduces  measurement statistics of the same observables for a state $\mathrm{F}(\rho)$ on the configuration $\rho\otimes\mathrm{F}(\rho)$, or equivalently $\rho$ on the configuration $\mathrm{F}(\rho)\otimes\rho$. Let
\begin{align}
\Pi^{\theta}_{{[a,b,c]}}:=\frac{1}{2}\Pi^{\prime~\theta}_{{[a,b,c]}}+\frac{1}{2}(\mathrm{F}\otimes\mathrm{F})\left(\Pi^{\prime\prime~\theta}_{{[a,b,c]}}\right).
\end{align}
Applying $\mathrm{F}\otimes\mathrm{F}$ on both sides we get
\begin{align}
(\mathrm{F}\otimes\mathrm{F})\left(\Pi^{\theta}_{{[a,b,c]}}\right)&=\frac{1}{2}(\mathrm{F}\otimes\mathrm{F})\left(\Pi^{\prime~\theta}_{{[a,b,c]}}\right)+\frac{1}{2}\Pi^{\prime\prime~\theta}_{{[a,b,c]}}\nonumber\\
&=\frac{1}{2}(\mathrm{F}\otimes\mathrm{F})\left(\Pi^{\prime\prime~\theta}_{{[-a,-b,-c]}}\right)+\frac{1}{2}\Pi^{\prime~\theta}_{{[-a,-b,-c]}}=\Pi^{\theta}_{{[-a,-b,-c]}}.
\end{align}

\noindent{\bf (3)} Let the POVM $\left\{\Pi^{\prime~\theta}_{{[a,b,c]}}\right\}$ reproduces measurement statistics of $\left\{\sigma_{\hat{n}_i}\right\}_{i=1}^3$ for a state $\rho$ on the configuration $\rho\otimes\mathrm{F}(\rho)$. Manifestly, the POVM $\left\{\Pi^{\prime\prime~\theta}_{{[a,b,c]}}:=\Pi^{\prime~\theta}_{{[b,a,c]}}\right\}$ reproduces  measurement statistics of the same observables for a state $\mathrm{R_{XY}}(\rho)$ on the configuration $\rho\otimes\mathrm{F}(\rho)$, or equivalently $\rho$ on the configuration $(\mathrm{R_{XY}}\otimes\mathrm{R_{XY}})(\rho\otimes\mathrm{F}(\rho))$. Let
\begin{align}
\Pi^{\theta}_{{[a,b,c]}}:=\frac{1}{2}\Pi^{\prime~\theta}_{{[a,b,c]}}+\frac{1}{2}(\mathrm{R_{XY}}\otimes\mathrm{R_{XY}})\left(\Pi^{\prime\prime~\theta}_{{[a,b,c]}}\right).
\end{align}
Applying $\mathrm{R_{XY}}\otimes\mathrm{R_{XY}}$ on both sides we get
\begin{align}
(\mathrm{R_{XY}}\otimes\mathrm{R_{XY}})\left(\Pi^{\theta}_{{[a,b,c]}}\right)&=\frac{1}{2}(\mathrm{R_{XY}}\otimes\mathrm{R_{XY}})\left(\Pi^{\prime~\theta}_{{[a,b,c]}}\right)+\frac{1}{2}\Pi^{\prime\prime~\theta}_{{[a,b,c]}}\nonumber\\
&=\frac{1}{2}(\mathrm{R_{XY}}\otimes\mathrm{R_{XY}})\left(\Pi^{\prime\prime~\theta}_{{[b,a,c]}}\right)+\frac{1}{2}\Pi^{\prime~\theta}_{{[b,a,c]}}=\Pi^{\theta}_{{[b,a,c]}}.
\end{align}
Similarly, $\Pi^{\theta}_{[a,c,b]}=\mathrm{R_{YZ}}\otimes\mathrm{R_{YZ}}\left(\Pi^{\theta}_{[a,b,c]}\right)$and $\Pi^{\theta}_{[c,b,a]}=\mathrm{R_{ZX}}\otimes\mathrm{R_{ZX}}\left(\Pi^{\theta}_{[a,b,c]}\right)$.\\
This completes the proof of Lemma \ref{lem1}. 
\end{proof}

\noindent From statements {\bf (2)} and {\bf (3)} of Lemma \ref{lem1} one can see that to specify $\mathcal{G}^{_{\uparrow\hspace{-.05cm}\downarrow,\theta}}$ it is sufficient to specify the operators $\Pi^{\theta}_{[1,1,1]}$ and $\Pi^{\theta}_{[1,1,-1]}$ only, which according to statement {\bf (1)} of Lemma \ref{lem1} are Rank-1 operators. Assuming $\Pi^{\theta}_{[1,1,1]}\propto\ket{\chi}\bra{\chi}$ and $\Pi^{\theta}_{[1,1,-1]}\propto\ket{\eta}\bra{\eta}$, the unnormalized states $\ket{\chi}$ and $\ket{\eta}$ are specified in the following Lemma. 
\begin{manuallemma}{2S}\label{lem2}
$\ket{\chi}=\id\otimes K_{\chi}\ket{\psi^-}$ and $\ket{\eta}=\id\otimes K_{\eta}\ket{\psi^-}$, where $K_{\chi}=\id+u\left(\sum_{i=1}^3\mathrm{P}^{-1}_{\hat{n}_i}\right)$ and $K_{\eta}=\id+v\left(\sum_{i=1}^2\mathrm{P}^{-1}_{\hat{n}_i}\right)+w\mathrm{P}^{1}_{\hat{n}_3}$, with
\begin{align*}
u=\frac{4}{3(\cos2\theta-1)},~~~v=\frac{-4}{(5+3\cos2\theta)},~~~w=\frac{-4(7+9\cos2\theta)}{(5+3\cos2\theta)^2}~.
\end{align*}
Here, $\ket{\psi^-}$ denotes the two-qubit singlet state. 
\end{manuallemma}
\begin{proof}
From CJ vector isomorphism we can write $\ket{\chi}=\id\otimes K_{\chi}\ket{\psi^-}$.  Since the operators $\left\{\id,\mathrm{P}^{-1}_{\hat{n}_1},\mathrm{P}^{-1}_{\hat{n}_2},\mathrm{P}^{-1}_{\hat{n}_3}\right\}$ forms a basis of $\mathcal{L}(\mathbb{C}^2)$ we can express $K_{\chi}$ (upto normalization) as
\begin{align}
K_{\chi}=\id+\sum_{i=1 }^3u_i\mathrm{P}^{-1}_{\hat{n}_i}.  
\end{align}
Expressing the singlet state as $\ket{\psi^-}\propto\ket{\psi^{-1}_{n_{i}}\psi^{1}_{n_{i}}}-\ket{\psi^{1}_{n_{i}}\psi^{-1}_{n_{i}}}$, from Eq.(\ref{ortho}) we have 
\begin{align*}
\bra{\psi^{-1}_{n_{i}}}\otimes\bra{\psi^{1}_{n_{i}}}\id\otimes K_{\chi}\ket{\psi^-}=0\quad \Rightarrow \bra{\psi^{1}_{n_{i}}}K_{\chi}\ket{\psi^{1}_{n_{i}}}=0,~\forall~i.
\end{align*}
This uniquely solves $u_i$'s and yields
\begin{align}
u_1=u_2=u_3:=u=\frac{4}{3(\cos2\theta-1)}.
\end{align}
Similar analysis for $K_{\eta}$ fixes the values of $v$ and $w$, and completes the proof.  
\end{proof}
\noindent We have almost identified the desired POVM apart from the normalization factors appearing with its Rank-1 effects. These normalization factors are dependent on the angle $\theta$ and put constraints on existence of the parent POVM as established in our next Theorem. 
\begin{manualtheorem}{2S}\label{theo3s}
The set of observables $\left\{\sigma_{\hat{n}_i}\right\}_{i=1}^3$ is jointly measurable on 2-copy {\footnotesize $\uparrow\hspace{-.05cm}\downarrow$} configuration if and only if $\theta\in[0,\cos^{-1}(1/3)]$.
\end{manualtheorem}
\begin{proof}
Assuming normalization for $\Pi_{[1,1,1]}$ and $\Pi_{[1,1,-1]}$ as $N$ and $M$ respectively and using Lemma~\ref{lem2}, we get 
\begin{subequations}
\begin{align}
\Pi_{[1,1,1]}&=N\ket{\chi}\bra{\chi}\nonumber\\
&= N\Big(p~\id^{\otimes 2} + q~(\dcb{X,\id}~+ \dcb{Y,\id}~+ \dcb{Z,\id}~)\nonumber\\
&~~~~+ r~(\dacb{X,Y}~ + \dacb{Y,Z}~ + \dacb{Z,X}~)+s~\big(X^{\otimes 2}+Y^{\otimes 2}+Z^{\otimes 2}\big)~\Big)~,\\
\Pi_{[1,1,-1]}&=M\ket{\eta}\bra{\eta}\nonumber\\
&= M\Big(p_1\id^{\otimes 2} + q_1(\dcb{X,\id}~+ \dcb{Y,\id}~)~+ q_2~\dcb{Z,\id}~\nonumber\\
&~~~+ r_1~( \dacb{Y,Z}~ + \dacb{Z,X}~)+ r_2\dacb{X,Y}~+ s_1\big(X^{\otimes 2} + Y^{\otimes 2}\big)~+s_2 Z^{\otimes 2}~\Big),
\end{align}
\end{subequations}
where,
\begin{subequations}
\begin{align}
\begin{rcases}
p = 1+\frac{3}{2} u \left(2+u \left(2+2 \alpha  \beta +\beta ^2\right)\right),\\
q = \frac{1}{2} u (2+ 3 u) (\alpha +2 \beta ),\\
r = -\frac{1}{2} u^2 (1+2 \beta  (2 \alpha +\beta )), \\
s = -1+\frac{1}{2} u \left(-6+u \left(-4+2 \alpha  \beta +\beta ^2\right)\right)
\end{rcases},
\end{align}
\begin{align}
\begin{rcases}
p_1 = \frac{1}{2} \left(2+4v+w (2+w)-2 v w \left(-1+2 \alpha  \beta +\beta ^2\right)+v^2 \left(3+ 2 \alpha  \beta +\beta ^2\right)\right),\\
q_1= \frac{1}{2} (2+ 2 v+w) (-w \beta +v (\alpha +\beta )),\\
q_2= \frac{1}{2} (2+ 2 v+w) (-w \alpha+2 v \beta ),\\
r_1= \frac{1}{2} \left(-w^2 \alpha  \beta -2 v^2 \beta   (\alpha +\beta )+v (w+w\alpha  \beta)\right), \\
r_2= \frac{1}{2} \left(- w^2\beta ^2+2 v w\beta   (\alpha +\beta )+v^2 \left(-1-2 \alpha  \beta +\beta ^2\right)\right),\\
s_1= -(1+v) (1+v+w)-vw\alpha  \beta   -\frac{1}{2}  \left(-2 v^2+w^2\right)\beta ^2,\\
s_2= \frac{1}{2} \left(-2+v^2 \left(-1+2 \alpha  \beta -3 \beta ^2\right)-2 v \left(2+w+w\beta ^2\right)+w \left(-2+w\left(-1+2 \beta ^2\right)\right)\right)
\end{rcases}.
\end{align}
\end{subequations}
Now, the reproducibility condition of Proposition \ref{prop1s} for the parent POVM further demands
\begin{align}
\begin{rcases}
&Np+3Mp_1=\frac{1}{2},~~Nq+M(2q_1-q_2)=\frac{\alpha}{4},~~Nq+Mq_2=\frac{\beta}{4},\\
&\hspace{1cm}Nr+M(2r_1+r_2)=0,~~Ns+M(2s_1+s_2)=0.
\end{rcases}\label{nor}
\end{align}    
It turns out that the equations in (\ref{nor}) are linearly dependent and yield unique solution for $N$ and $M$ as below
\begin{subequations}
\begin{align}
N(\theta)&=\frac{(7+9\cos2\theta)\tan^4\theta}{128},\\
M(\theta)&=\frac{(5+3\cos2\theta)^4\tan\theta}{128~ (3\sin2\theta)^3}.
\end{align}
\end{subequations}
Note that, $N(\theta)\ge0$ for $\theta\in(0,\cos^{-1}(1/3)]$ and yield consistent parent POVM. However, $N(\theta)<0$ for $\theta\in(\cos^{-1}(1/3),\pi/2]$ (see Fig.~\ref{Figs2}), and hence no parent POVM exists. This complete the proof. 
\begin{figure}[t!]
\centering
\includegraphics[height=7cm,width=11cm]{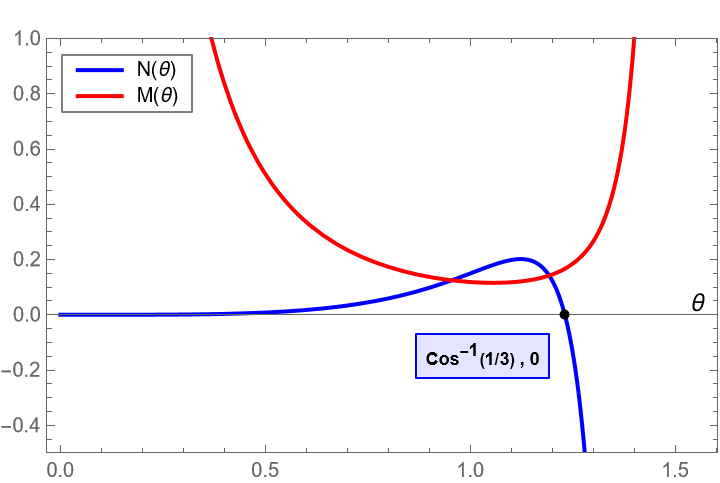}
\caption{(Color online) Plot of the normalization factors $N(\theta)$ and $M(\theta)$ with $\theta$. $N(\theta)<0$ for $\theta\in(\cos^{-1}(1/3),\pi/2]$. \vspace{0cm}} \label{Figs2}
\end{figure}
\end{proof}
\begin{manualremark}{3S}
Putting $\theta=\cos^{-1}(1/\sqrt{3})$ in Theorem \ref{theo3s} we obtain, 
\begin{align*}\begin{rcases}
N=M=1/8;~~p=p_1=1;~~q=q_1=r_1=1/2;~~\\
\hspace{1cm}r=q_2=r_2=-1/2;~~s=s_1=s_2=0~.
\end{rcases}
\end{align*}
This is the particular case of observables $\{X,Y,Z\}$ discussed in \textbf{Theorem 1} of the main manuscript. 
\end{manualremark}
\begin{manualremark}{4S}
Putting $\theta=\cos^{-1}(1/3)$ in Theorem \ref{theo3s} we obtain, 
\begin{align*}\begin{rcases}
N=0;~~M=1/6;~~p_1=1;~~q_1=1/\sqrt{3};~~q_2=-1/2\sqrt{3};\\
\hspace{1cm}r_1=s_2=1/3;~~r_2=-2/3,~~s_1=-1/6~.
\end{rcases}
\end{align*}
Accordingly we have $\Pi_{[1,1,1]}=0=\Pi_{[-1,-1,-1]}$, and the other six effects are given by
\begin{table}[h!]
\begin{tabular}{c c}
\large$\frac{1}{6}~\times~$ &
\begin{tabular}{c||c|c|c|c|c|c|c|c|c|c|}
\hline
& $\id^{\otimes2}$ & $\dcb{X,\id}$ & $\dcb{Y,\id}$ & $\dcb{Z,\id}$ & $\dacb{X,Y}$ & $\dacb{Y,Z}$ & $\dacb{Z,X}$ & $X^{\otimes2}$ & $Y^{\otimes2}$ & $Z^{\otimes2}$ \\ \hline\hline

$\Pi_{[1,1,-1]}$ & $1$ & $\tfrac{1}{\sqrt{3}}$ & $\tfrac{1}{\sqrt{3}}$ & $-\tfrac{1}{2\sqrt{3}}$ & $-\tfrac{2}{3}$ & $\tfrac{1}{3}$ & $\tfrac{1}{3}$ & $-\tfrac{1}{6}$ & $-\tfrac{1}{6}$ & $\tfrac{1}{3}$  \\ \hline

$\Pi_{[1,-1,1]}$ & $1$ & $\tfrac{1}{\sqrt{3}}$ & $-\tfrac{1}{2\sqrt{3}}$ & $\tfrac{1}{\sqrt{3}}$ & $\tfrac{1}{3}$ & $\tfrac{1}{3}$ & $-\tfrac{2}{3}$ & $-\tfrac{1}{6}$ & $\tfrac{1}{3}$ & $-\tfrac{1}{6}$  \\ \hline

 $\Pi_{[-1,1,1]}$ & $1$ & $-\tfrac{1}{2\sqrt{3}}$ & $\tfrac{1}{\sqrt{3}}$ & $\tfrac{1}{\sqrt{3}}$ & $\tfrac{1}{3}$ & $-\tfrac{2}{3}$ & $\tfrac{1}{3}$ & $\tfrac{1}{3}$ & $-\tfrac{1}{6}$ & $-\tfrac{1}{6}$  \\ \hline

 $\Pi_{[-1,-1,1]}$ & $1$ & $-\tfrac{1}{\sqrt{3}}$ & $-\tfrac{1}{\sqrt{3}}$ & $\tfrac{1}{2\sqrt{3}}$ & $-\tfrac{2}{3}$ & $\tfrac{1}{3}$ & $\tfrac{1}{3}$ & $-\tfrac{1}{6}$ & $-\tfrac{1}{6}$ & $\tfrac{1}{3}$  \\ \hline

$\Pi_{[-1,1,-1]}$ & $1$ & $-\tfrac{1}{\sqrt{3}}$ & $\tfrac{1}{2\sqrt{3}}$ & $-\tfrac{1}{\sqrt{3}}$ & $\tfrac{1}{3}$ & $\tfrac{1}{3}$ & $-\tfrac{2}{3}$ & $-\tfrac{1}{6}$ & $\tfrac{1}{3}$ & $-\tfrac{1}{6}$  \\ \hline

 $\Pi_{[1,-1,-1]}$ & $1$ & $\tfrac{1}{2\sqrt{3}}$ & $-\tfrac{1}{\sqrt{3}}$ & $-\tfrac{1}{\sqrt{3}}$ & $\tfrac{1}{3}$ & $-\tfrac{2}{3}$ & $\tfrac{1}{3}$ & $\tfrac{1}{3}$ & $-\tfrac{1}{6}$ & $-\tfrac{1}{6}$  \\ \hline
\end{tabular}
\end{tabular}
\caption{The coefficients of the effects are show in the Table. Each values in the table are multiplied by $1/6$, as indicated at the right side of the Table.}
\end{table}

\noindent Note that,
\begin{subequations}
\begin{align}
\tr\left[\left(\Pi_{[1,1,-1]}+\Pi_{[1,-1,1]}+\Pi_{[1,-1,-1]}\right)\Big(\rho\otimes\mathrm{F}(\rho)\Big)\right]=\tr\left[\mathrm{P}^{+1}_{\hat{m}_1}\rho\right],\\
\tr\left[\left(\Pi_{[1,1,-1]}+\Pi_{[-1,1,1]}+\Pi_{[-1,1,-1]}\right)\Big(\rho\otimes\mathrm{F}(\rho)\Big)\right]=\tr\left[\mathrm{P}^{+1}_{\hat{m}_2}\rho\right],\\
\tr\left[\left(\Pi_{[1,-1,1]}+\Pi_{[-1,1,1]}+\Pi_{[-1,-1,1]}\right)\Big(\rho\otimes\mathrm{F}(\rho)\Big)\right]=\tr\left[\mathrm{P}^{+1}_{\hat{m}_3}\rho\right],
\end{align}
\end{subequations}
where
\begin{align}
\hat{m}_1=\tfrac{1}{3\sqrt{3}}(5,-1,-1)^{\mathrm{T}},~~\hat{m}_2=\tfrac{1}{3\sqrt{3}}(-1,5,-1)^{\mathrm{T}},~~\hat{m}_3=\tfrac{1}{3\sqrt{3}}(-1,-1,5)^{\mathrm{T}}.
\end{align}
Furthermore, we also have 
\begin{align}
&\tr\left[\left(\Pi_{[-1,-1,1]}+\Pi_{[-1,1,-1]}+\Pi_{[1,-1,-1]}\right)\Big(\rho\otimes\mathrm{F}(\rho)\Big)\right]=\tr\left[\mathrm{P}^{+1}_{\hat{m}_0}\rho\right],\\
&\hspace{4cm}\text{where}~\hat{m}_0=\tfrac{1}{\sqrt{3}}(-1,-1,-1)^{\mathrm{T}}.
\end{align}

\noindent Also note that 
\begin{align}
\hat{m}_i\cdot\hat{m}_{j\neq i}=-\tfrac{1}{3},~\forall~i,j\in\{0,1,2,3\}~. 
\end{align}
Thus, the set of vectors $\{\hat{m}_r\}_{r=0}^3$ are related with the vectors $\{\hat{n}_r\}_{r=0}^3$ in \textbf{Theorem \ref{theo2}} of the main manuscript as follows
\begin{align}
\hat{m}_0=-\hat{n}_0,~~\&~~\hat{m}_r=\mathcal{R}^{\hat{m}_0}_\varphi(-\hat{n}_r),~~\text{for}~~ r\in\{1,2,3\},    
\end{align}
where $\mathcal{R}^{\hat{m}_0}_\varphi$ denotes rotation about $\hat{m}_0$ by an angle $\varphi=\cos^{-1}(7/9)$.
\end{manualremark}


\begin{thebibliography}{74}
\bibitem [{Sel({\natexlab{a}})}]{Self1}%
  \BibitemOpen
  \bibfield  {title} {\bibinfo {title} {The set of effect operators $\mathcal{E}(\mathcal{H})$ forms a convex and compact subset of the cone of positive operators $\mathcal{P}(\mathcal{H})$, which itself is a convex subset of the space of linear operators $\mathcal{L}(\mathcal{H})$},\ }\href@noop {} {\  ({\natexlab{a}})}\BibitemShut {NoStop}%
\bibitem [{Sel({\natexlab{c}})}]{Self4}%
  \BibitemOpen
  \bibfield  {title} {\bibinfo {title} {The dual map ${\Lambda}^\star$ of ${\Lambda}$ is defined via the relation $\tr[{A}~ {\Lambda}({B})] = \tr[{\Lambda}^\star({A})~ {B}]$ for all ${A},{B} \in \mathcal{L}(\mathcal{H})$},\ }\href@noop {} {\  ({\natexlab{c}})}\BibitemShut {NoStop}%
\bibitem [{Sel({\natexlab{d}})}]{Self5}%
  \BibitemOpen
  \bibfield  {title} {\bibinfo {title} {In the original version {VAA} have considered $\ket{\phi^+}$ instead of $\ket{\psi^-}$, and hence their protocol is local unitray equivalent to the protocol discussed here. {M}ore particularly, the original version can be reproduced by applying our analysis to the configuration $\rho_{\vec{m}}\otimes\rho^\star_{\vec{m}}$ instead of $\rho_{\vec{m}}\otimes\rho_{-\vec{m}}$, where $\rho^\star_{\vec{m}}=\tfrac{1}{2}(\id+m_x\sigma_x-m_y\sigma_y+m_z\sigma_z)$},\ }\href@noop {} {\  ({\natexlab{d}})}\BibitemShut {NoStop}%
\end{thebibliography}
\end{document}